\documentclass{CVM}

\CVMsetup{
type      = {Research Article},
doi       = {s41095-0xx-xxxx-x},
title     = {MusicFace: Music-driven Expressive Singing Face Synthesis},
author    = {Pengfei Liu$^{1}$, Wenjin Deng$^{1}$, Hengda Li$^{1}$, Jintai Wang$^{1}$, Yinglin Zheng$^{1}$, Yiwei Ding$^{1}$, Xiaohu Guo$^{2}$, and Ming Zeng$^{1}$\cor{}
},
runauthor = {P. Liu, W. Deng, H. Li, \emph{et al.}},
abstract  = {
  It is still an interesting and challenging problem to synthesize a vivid and realistic singing face driven by music signal. In this paper, we present a method for this task with natural motions of the lip, facial expression, head pose, and eye states. Due to the coupling of the mixed information of human voice and background music in common signals of music audio, we design a decouple-and-fuse strategy to tackle the challenge. We first decompose the input music audio into human voice stream and background music stream. Due to the implicit and complicated correlation between the two-stream input signals and the dynamics of the facial expressions, head motions and eye states, we model their relationship with an attention scheme, where the effects of the two streams are fused seamlessly. Furthermore, to improve the expressiveness of the generated results, we propose to decompose head movements generation into speed generation and direction generation, and decompose eye states generation into the short-time eye blinking generation and the long-time eye closing generation to model them separately. We also build a novel SingingFace Dataset to support the training and evaluation of this task, and to facilitate future works on this topic. Extensive experiments and user study show that our proposed method is capable of synthesizing vivid singing face, which is better than state-of-the-art methods qualitatively and quantitatively.
},
keywords  = {Music-driven face generation, singing face, generative adversarial network.},
copyright = {The Author(s)},
}

\begin{document}

\maketitle

    \begin{figure}[b] \vskip -4mm
    \small\renewcommand\arraystretch{1.3}
        \begin{tabular}{p{80.5mm}} \toprule \\ \end{tabular}
        \vskip -4.5mm \noindent \setlength{\tabcolsep}{1pt}
        \begin{tabular}{p{3.5mm}p{80mm}}
    $1\quad $ & School of Informatics, Xiamen University, Xiamen, 361000, China. E-mail: P. Liu, liupengfei@stu.xmu.edu.cn; W. Deng, dengwenjin@stu.xmu.edu.cn; H. Li, lihengda@stu.xmu.edu.cn; J. Wang, jintaiwang@stu.xmu.edu.cn; Y. Zheng, zhengyinglin.stu.xmu.edu.cn; Y. Ding, dingyiwei@stu.xmu.edu.cn;  M. Zeng, zengming@xmu.edu.cn\cor{}, the corresponding author.\\
    $2\quad $ & Department of Computer Science, The University of Texas at Dallas, Richardson, 75080-3021, Texas, US. E-mail: xguo@utdallas.edu.\\

    \end{tabular} \vspace {-3mm}
    \end{figure}

\section{Introduction}\label{sec:introduction}
With the advancement of computer vision and computer graphics, synthesizing vivid and realistic dynamic face is becoming possible and has been attracting more and more attention from CV/CG communities. Recent progresses~\cite{VOCA2019,suwajanakorn2017synthesizing,chen2020talkinghead,yi2020audio,c31,c34,c35, yi2020audio} show the great potential of this topic in a variety of applications, such as human-computer interaction~\cite{c41,c42}, video making~\cite{c36,c37,c38,c46}, and news anchor composition\cite{c39,c40}, \emph{etc.}

Despite the recent progresses of dynamic face synthesis~\cite{VOCA2019,suwajanakorn2017synthesizing,chen2020talkinghead,yi2020audio,c31,c34,c35,zhou2021pose,c45} and its potential applications~\cite{c41,c42,c36,c37,c38,c46,c39,c40}, 

it is still an open problem regarding how to synthesize a vivid face as expressive as possible. 

\begin{figure}
  \includegraphics[width=0.98\linewidth]{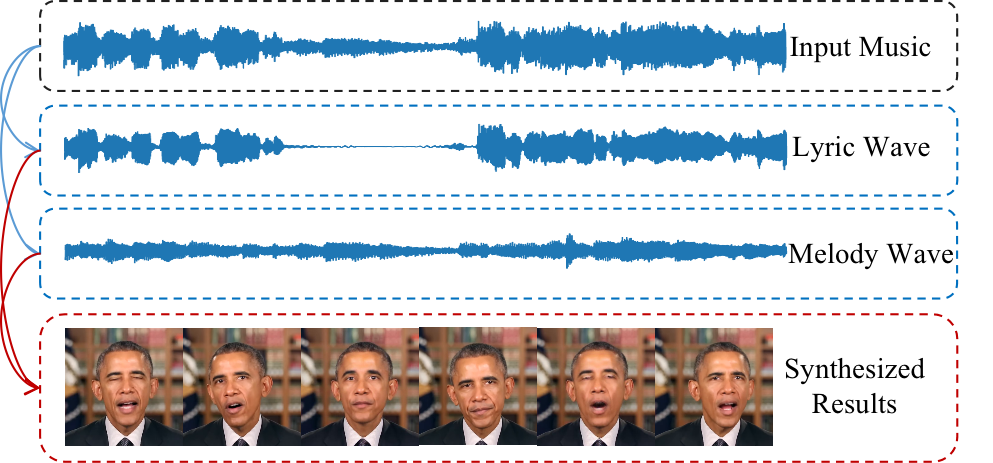}
  \caption{\textbf{Problem description.} Our goal is to synthesize a vivid dynamic singing face coherent with the input music audio, which is mixed with human voice and background music.}
  \label{fig:problem_description}
\end{figure}

Existing work in the literature focuses on generating coherent dynamics of faces according to input speech audio~\cite{c20,c23,VOCA2019,suwajanakorn2017synthesizing,wav2lip,NeuralVoicePuppetry,c28,chen2020talkinghead,yi2020audio,c31}. However, in many emotional scenarios, it is required that the head synthesis is driven by a composite audio which is coupled with not only speech but also other signals, \emph{e.g.}, ~the music audio contains both human voice and background music signals. Therefore, in this paper, we investigate the problem of synthesizing a vivid dynamic face which is not only in-sync but also delivers coherent facial dynamics with the input music audio, as is illustrated in Fig.~\ref{fig:problem_description}. This is a non-trivial task, which can not be handled directly by existing methods. This is because common music audios are mixed by coupled human voice and background music signals, while most of the existing methods are designed for synthesizing face according to only the human speech signals, which will lead to undesired results due to the entanglement of different audio signals. 

To tackle this challenge, we investigate the implicit correlation between the input signals and the facial dynamics. 
 We treat the input music audio as a mixed signal which includes a human voice signal and a background music signal. According to previous work~\cite{suwajanakorn2017synthesizing,wav2lip,NeuralVoicePuppetry,c28,chen2020talkinghead,yi2020audio,c31} and our observation, we argue that the lip movement is majorly related to the voice signal (also called the speech channel), while the head pose, facial expression, eye states relate to both the voice signal and background music signal. However, we would like to ask the questions: \emph{Are these subjective observations true?} and \emph{How much do the human voice and background music signals affect the face dynamics?} To answer these questions, we devise a decouple-and-fusion framework for this task. Firstly, we separate the input music audio into the human voice channel and the background music channel. Then we dynamically fuse these two separated signals in a feature selection fashion by introducing a \emph{Attention-based Modulator}. The \emph{Attention-based Modulator} modulates and balances the two signals for the downstream generators of facial expressions, head motions, and eye states.  

In the singing scenarios, the motions of the head and eyes are usually emotional and dramatic, which raises challenges for generators to learn the more diverse and expressive motions as compared with the previous talking scenarios. We propose two ingredients to improve the expressiveness of the synthesis result. For the movement of the head, we propose to learn the rhythm of head motion that is decoupled from the absolute moving velocity, thus factoring off the ambiguity of the mapping between audio and head movement. For the eye states, we propose to synthesize both eye blinking and long-time eye closing states, which delivers much more expressiveness as compared with previous methods. 

Besides, to learn the complex and implicit relationship between the music audio and face dynamics, we build a SingingFace Dataset from our recordings. The dataset contains over 600 singing videos with synchronous music audio. 
To our best knowledge, this is the first dataset regarding face dynamics and music audio. We believe it will promote future research on this topic. 

In summary, this paper is featured as follows: 

\begin{itemize}
\setlength{\itemsep}{0pt}
\setlength{\parsep}{0pt}
\setlength{\parskip}{0pt}
    \item This is the first framework for synthesizing a singing face video driven by the input music audio mixed with human voice and background music signals. In the framework, we introduce the \emph{Attention-based Modulator} to balance the effects of the two signals on the head movements, expressions, and eye states.  
    
    \item We propose to synthesize the speed and direction of head movements separately, instead of predicting head pose directly. The simple-yet-effective modification leads to more consistent head dynamics in line with music rhythm. Besides, we propose to decompose the eye states into eye blinking and long-time eye closing, which is much more realistic in singing scenarios. 
    
    \item We build the first dataset which contains expressive singing face videos with synchronous music audio, and make it public to facilitate future research on this topic. 
\end{itemize}

\section{Related Work}\label{Relatedwork}

\begin{figure*}[ht!p]
\begin{center}
\includegraphics[width=0.90\textwidth]{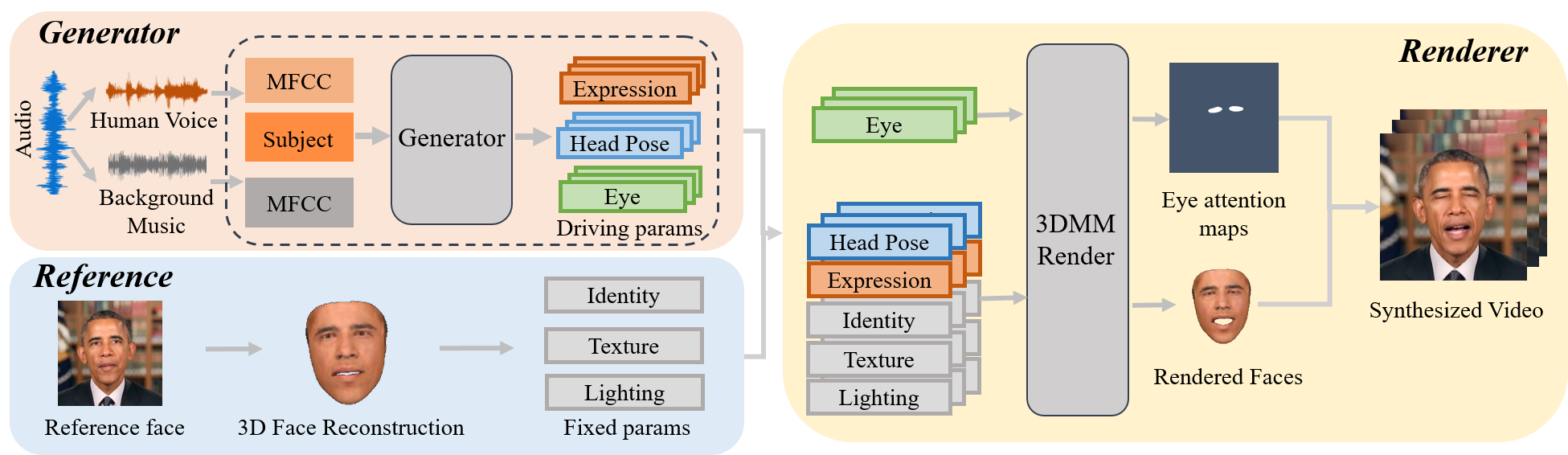}
\end{center}

\caption{\textbf{Framework overview.} Taking human voice and background music separated from music audio as input, the Generator module generates facial driving parameters~(expressions, head poses and eye states). 
Conditioned with fixed parameters~(identity, texture, lighting) extracted from a reference face image and the driving parameters, the Renderer module aims to synthesize a photo-realistic video. Specifically, eye state parameters are encoded into eye attention maps, and other parameters provide a 3D model guidance to render faces. Finally, an expressive and rhythmic singing face video is rendered by combining rendered faces with eye attention maps.}

\label{fig:schematic_overview}
\end{figure*}

\subsection{Audio-driven Talking Face Synthesis} 
Audio-driven face synthesis has been widely explored. Previous work~\cite{c19,c21,c22,c23,VOCA2019} focuses on establishing the mapping between facial motion factors and audio features. Brand~\cite{c19} uses a Hidden Markov Model~(HMM) to predict facial motions. Ezzat \textit{et al.}~\cite{c21} leverage an example-based method mapping phonemes to mouth shape and texture parameters in the Principle Component Analysis~(PCA) space. Wang \textit{et al.}~\cite{c22} attempt to model a mapping between Mel-Frequency Cepstral Coefficients~(MFCC) and PCA model parameters via an HMM approach. Benefiting from deep learning techniques, some works have been proposed to generate more diverse faces in sync with input audio. Shimba \textit{et al.}~\cite{c23} estimate active appearance model~(AAM) parameters with the Long Short-Term Memory~(LSTM) network. Cudeiro \textit{et al.}~\cite{VOCA2019} employ convolutions to encode speech and decode facial attributes to animate a 3D template.

Several methods~\cite{c20,wav2lip,NeuralVoicePuppetry, chen2019hierarchical,suwajanakorn2017synthesizing,Speech-Driven, fried2019text, zhou2019talking,yao2021iterative, guo2021ad, xie2021towards} merely synthesize facial region texture with lip-synced motions. Among the above approaches, a broad class of them generate identity-preserving face with static head pose using GANs~\cite{chen2019hierarchical, Speech-Driven, zhou2019talking}. Other methods synthesize lip-synced texture of mouth, then rewrite the mouth area of source frames according to the input audio~\cite{suwajanakorn2017synthesizing, NeuralVoicePuppetry, wav2lip, c20, xie2021towards} or text~\cite{fried2019text, yao2021iterative}. However, due to the dependence on the original video, they can only generate limited head poses. To address this problem, Chen \textit{et al.}~\cite{chen2020talkinghead}, Yi \textit{et al.}~\cite{yi2020audio} and Zhang \textit{et al.}~\cite{zhang20213d} estimate head movements from input audio. Most recently, Zhang \textit{et al.}~\cite{zhang2021facial}, Li \textit{et al.}~\cite{li2021write} and Guo \textit{et al.}~\cite{guo2021ad} synthesize photo-realistic 3D head with natural head poses and synchronized lip motions using popular neural rendering techniques. Wang \textit{et al.}~\cite{wang2021audio2head, wang2022one} even generate photo-realistic faces from one-shot reference image with natural motions.

\subsection{Music-driven Animation} 

Music-driven human pose animation has been studied for decades. Early work~\cite{c4, c5, c6} formulate the task as a template matching problem. Lee \textit{et al.}~\cite{c5} and Shiratori \textit{et al.}~\cite{c6} generate dance motion sequences with musical similarity based on manually defined audio features, while Cardle \textit{et al.}~\cite{c4} edit motions guided by musical features. Due to the limitations of capacity, these template matching approaches are not competent to generate diverse and natural dance motions.

With the great success of deep neural network, more researchers address the music-to-dance as a generation problem with learning-based techniques. Recent methods employ auto encoder-decoder~\cite{c17}, LSTM~\cite{c7,c8,c9,c10,c11}, GAN~\cite{c12,c13}, and Transformer~\cite{c14,c15,c16}. Even though some work~\cite{c12,c18} apply action units to further explore the correlations between pose and music, it is still challenging to generate diverse, rhythmic and expressive dance motion. 

It is interesting to note that music-driven singing face synthesis remains a rarely studied open problem. Song2Face~\cite{iwase2020song2face} is the only one designed for singing scenarios up to now to the best of our knowledge. However, it operates on plain human singing voice, only working well without the disturbance of background music. 

Synthesizing expressive singing faces from mixed music signals is more challenging and difficult in three aspects. Firstly, singing voice and background music are entangled together, making it difficult for models to extract phonemes related information, and further leading to inaccurate lip motions. In addition, the relative contributions of different driven sources change over time and are even interconnected with each other. Finally, the model should consider multiple downstream generation task at the same time to make the result look natural and realistic overall. To solve the above issues, this paper propose a decouple-and-fuse framework, that can generate realistic and rhythmical facial dynamics from mixed music wave. Therefore, the paper will open novel research directions in the domain of music-guided person synthesis.

\section{Methodology}

\begin{figure*}[ht!p]
\begin{center}
\includegraphics[width=0.95\textwidth,height=7.5cm]{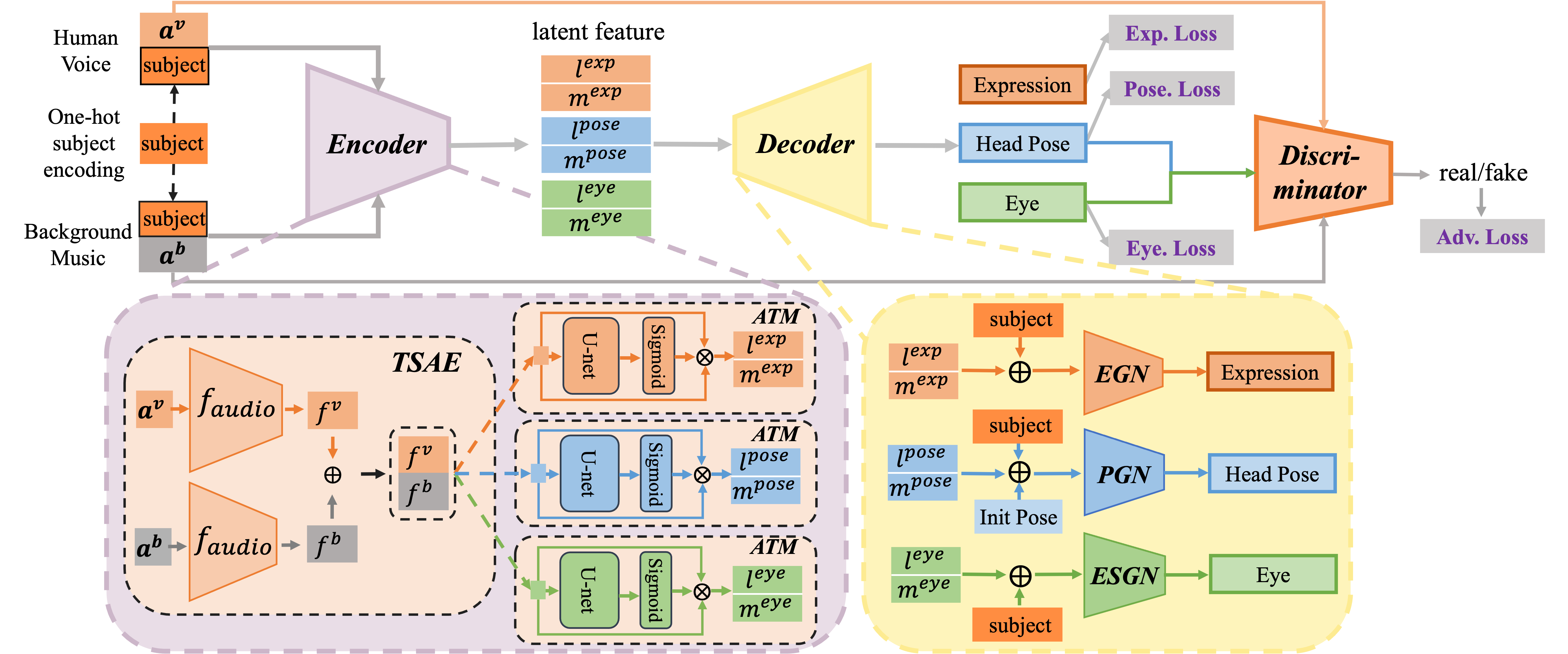}
\end{center}

\caption{\textbf{The Architecture of Our Generator.} Our generator contains an Encoder and a Decoder. The Encoder consists of a Two-stream Audio Encoder~(TSAE) and an Attention-based Modulator~(ATM). The Decoder contains three downstream generators, including Expression Generation Network~(EGN), Pose Generation Network~(PGN), and Eye State Generation Network~(ESGN).
}

\label{fig:training_phase}
\end{figure*}

\subsection{Problem Definition}

In previous researches~\cite{chen2019hierarchical,zhou2021pose,yi2020audio,zhang2021facial,sinha2020identitypreserving,zhou2020makelttalk}, given a piece of speech audio $\mathcal A$ and a short reference video (or a single face image) $\mathcal V$, the ultimate goal is to generate a realistic talking face video $\mathcal S$ synchronized with the input audio $\mathcal A$, which can be represented as:

\begin{equation}
\begin{split}
\mathcal F_{exp}, \mathcal F_{pose}, \mathcal F_{eye}=\mathbf{G}(\mathbf{E}(\mathcal A)), \\
\mathcal S=\mathbf{R}(\mathcal F_{exp}, \mathcal F_{pose}, \mathcal F_{eye}, \mathcal V),
\end{split}
\end{equation}
where $\mathcal F_{exp}$, $\mathcal F_{pose}$, $\mathcal F_{eye}$ denote the facial expression, head pose, and eye state parameters synthesized by a generator $\mathbf G$, respectively. $\mathbf E$ refers to an audio feature extractor and $\mathbf R$ denotes a rendering network synthesizing photo-realistic images.

However, directly predicting driving parameters from audio is not up to music scenarios due to the complicated mutual influences between human voice containing 
lyric information and background music containing melody information. We propose a decouple-and-fuse strategy to tackle the above problem, which firstly adopts an audio source separation model $\mathbf O$ to decompose music into human voice $\mathcal A^v$ and background music $\mathcal A^b$ , then gets encoded lyric feature $\mathcal L$ and melody feature $\mathcal M$ respectively using an attention-assisted two-stream encoder $\mathbf{E}$. It encodes lyric and melody separately, and modifies the relative contribution of the two encoded features on the generation process through an attention mechanism. Finally a generator $\mathbf G$ is employed to generate the driving parameters of a singing face video $\mathcal S$ from the decoupled lyric feature and melody feature. The full pipeline can be formulated as follows:

\begin{equation}
    \begin{split}
\mathcal A^v, \mathcal A^b &= \mathbf O(\mathcal A),\\
\mathcal L, \mathcal M &= \mathbf E(\mathcal A^v, \mathcal A^b),\\
\mathcal F_{exp}, \mathcal F_{pose}, \mathcal F_{eye} &= \mathbf G(\mathcal L, \mathcal M),\\ 
\mathcal S &= \mathbf R(\mathcal F_{exp}, \mathcal F_{pose}, \mathcal F_{eye}, \mathcal V).
    \end{split}
\end{equation}

As illustrated in Fig.~\ref{fig:schematic_overview}, our overall framework contains three components: 1) a driving parameter \textbf{generator} to translate music audio to facial expression, head pose, and eye states, 2) a \textbf{reference} module extracting fixed parameters such as face identity given a human face, 3) and a \textbf{renderer} to synthesize photo-realistic frames conditioned on above parameters. We employ a conditional-GAN-based method as our renderer, which is of the same architecture as~\cite{zhang2021facial}. To enhance the expressiveness of singing faces, the generator $\mathbf{G}$ is designed as the following encoder-decoder architecture as is shown in Fig.~\ref{fig:training_phase}. The Encoder~(Sec.~\ref{sec:encoder}) consists of a Two-stream Audio Encoder~(TSAE) to encode lyric and melody separately and an Attention-based Modulator~(ATM) to balance the contribution of different audio features. The Decoder~(Sec.~\ref{sec:decoder}) contains three downstream generators, including Expression Generation Network~(EGN) for the generation of facial expression parameters, Pose Generation Network~(PGN) for the generation of head pose dynamics, and Eye State Generation Network~(ESGN) for the generation of eye state parameters. In the next subsections, we will introduce the five essential parts respectively and provide the corresponding learning objective and training strategy.

\subsection{Encoder}
\label{sec:encoder}

As mentioned above, lyric and melody entangled in the original music wave show a complicated relationship in guiding the generation process of human face dynamics, making it difficult for the generation network to synthesize vivid face dynamics directly from plain music features. To tackle the problem, we employ a decouple-and-fuse strategy. Specifically, using a state-of-the-art audio source separation model Spleeter~\cite{spleeter2020}, we decompose the original music into human voice and background music. Then we encode lyric from human voice and melody from background music separately using a two-steam audio encoder. Finally, we adjust them with attention-based modulators to distribute the relative contribution of lyric and melody for each specific generation task.

\subsubsection{Audio Feature Extraction}

Taking the separated audio wave (human voice or background music) sampled at 16KHz of $T$ seconds as input, we extract mel-frequency cepstral coefficients~(MFCC) and their first derivatives with 25ms window size and 10ms window step, resulting in 26-D audio features of 100 frames per second. Furthermore, in order to incorporate temporal information and match the frequency of video frames~(30 fps), the feature sequence are converted to overlapping windows of size 39~(corresponding to 390ms ) at 30 fps. Therefore, the output feature is a three-dimensional array with the size $(30 \times T, 39, 26)$.

\subsubsection{Two-stream Audio Encoder~(TSAE)}
Given the separated human voice feature $\mathcal A^v$ and background music feature $\mathcal A^b$, we adopt a Two-stream Audio Encoder~(TSAE) that consists of two networks $\text{AE}^{v}$ and $\text{AE}^{b}$ to encode the MFCC features of human voice $a^v_t$ and background music $a^b_t$, separately:

\begin{equation}
\begin{split}
f^{v}_t &= \text{AE}^{v}(a^{v}_t),\\
f^{b}_t &= \text{AE}^{b}(a^{b}_t),
\end{split}
\end{equation}
where $\text{AE}^v$ and $\text{AE}^b$ are 1D temporal convolutional neural networks with residual blocks sharing the same network structure , and $f^v_t$, $f^b_t$ indicate the encoded audio features. The subscript $t$ indicates time step, and the superscripts $v$ and $b$ indicate human voice and background music, respectively. The encoded audio features of the full audio sequence $\mathbf{f}^v$ and $\mathbf{f}^b$ are obtained after stacking the audio features of each time step.

\subsubsection{Attention-based Modulator~(ATM)}

For a specific downstream generation task, the relative contributions of features representing different specific semantic information change over time and are even interconnected with each other. For example, image the head pose dynamics of a person singing a line of a song. He will prepare to vocalize, then sing, and shut his mouth finally. In the first and third stages, he rotates his head rhythmically dominated by melody. But when he vocalizes, melody in background music and lyric in human voice influence his head movements together. So the dominant source changes over time and even becomes ambiguous during vocalization, making the generation task difficult.

Therefore, in order to generate vivid human face movements, we introduce a channel attention mechanism similar to the attention mechanism proposed in~\cite{hu2018squeeze} to determine the relative contribution between lyric and melody on the generation result. The only difference is that, to consider the long-time dependence between the audio features of different time steps, we select a temporal U-net to generate attention weights instead of using a simple multi layer perceptron~(MLP) network. Specifically, given the separately encoded audio features, we employ an Attention-based Modulator(ATM) for each generation task to estimate an attention weight of each feature map in embedding feature $\mathbf{f^v}$ and $\mathbf{f^b}$ to adjust the relative importance between them:

\begin{equation}
\begin{split}
\mathbf{att}&=\sigma(\textbf{U-net}(\mathbf{f^v} \oplus \mathbf{f^b})),\\
[\mathbf{l}, \mathbf{m}]&=\mathbf{ATM}(\mathbf{f^v} \oplus \mathbf{f^b})\\
&=\mathbf{att}\odot(\mathbf{f^v} \oplus \mathbf{f^b}),\\
\end{split}
\end{equation}
where $\mathbf{l}$ and $\mathbf{m}$ denote the final output embedding of lyric and melody features for the full audio sequence respectively, $\oplus$ represents the concatenate operation on the feature channel dimension, and $\odot$ indicates the element-wise product. $\mathbf{ATM}$ indicates the Attention-based Modulator implemented using an temporal u-net network $\textbf{U-net}$ and $\sigma$ represents the sigmoid activation function.

As shown in Fig.~\ref{fig:training_phase}, we employ one \textbf{ATM} to learn the optimal attention weight for each downstream task. Specifically, we apply a total of three \textbf{ATM}s on $\mathbf{f^v}$ and $\mathbf{v^b}$, to get $\mathbf{l^{exp}}$ and $\mathbf{m^{exp}}$ for expression generation task, $\mathbf{l^{pose}}$ and $\mathbf{m^{pose}}$ for head pose generation task, and $\mathbf{l^{eye}}$ and $\mathbf{m^{eye}}$ for eye state generation task, respectively. 

\subsubsection{Subject Style Embedding}

Our TSAE, EGN, PGN, and ESGN are conditioned on the subject code to learn subject-specific styles, adopting a similar strategy in \cite{VOCA2019}, which encodes each subject in the dataset using a one-hot subject encoding. At training stage, the subject encoding is concatenated to each input MFCC feature $a^v_t$ and $a^b_t$, and also concatenated to the final output $l_t$ and $m_t$ of the ATM.

\subsection{Decoder}
\label{sec:decoder}
\subsubsection{Expression Generation Network}

We employ a simple MLP consisting of two fully connected layers and one ReLU activation layer to regress facial expression~(including lip motion) parameters from the encoded lyric and melody features. The process can be formulated as:

\begin{equation}
    \hat{f}_t=\varphi_{exp}(l^{exp}_t \oplus m^{exp}_t),
\end{equation}
where ${\hat{f}_t}$ denotes the predicted facial expression parameter at time step $t$ and $\varphi_{exp}$ means the MLP for expression generation.

\subsubsection{Pose Generation Network}
\begin{figure}
    \centering
    \includegraphics[width=0.5\textwidth]{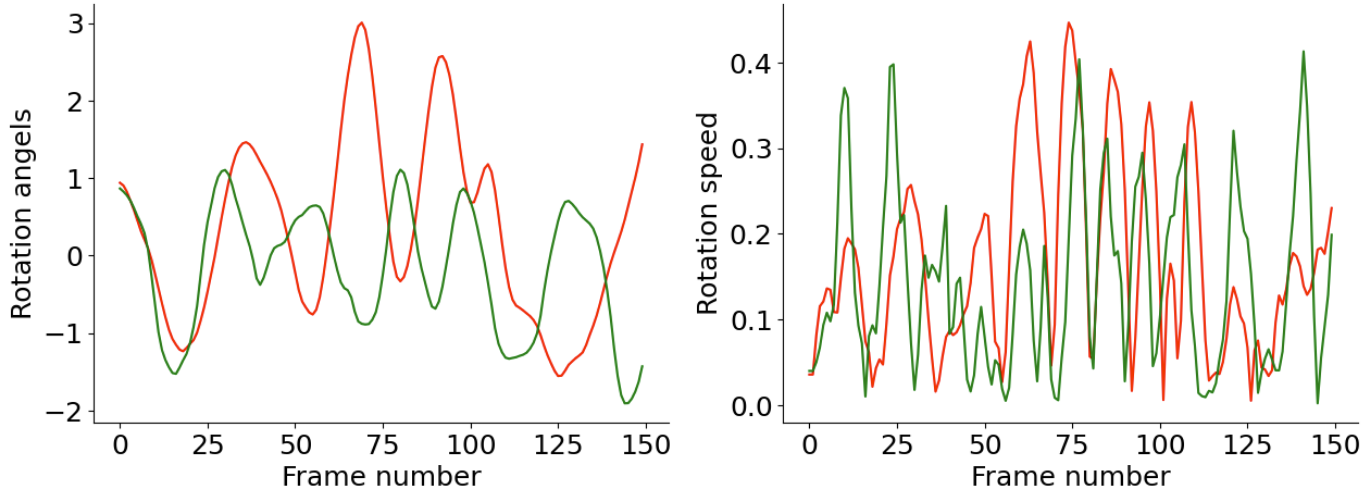}
    \caption{\textbf{The Euler angle ($R_y$) dynamics of a person singing the same song twice.} It shows although the head may rotate in opposite direction, the speed still keeps similar. This observation is also valid for other head pose parameters.}
    \label{fig:head_pose_dynamics}
\end{figure}

Traditional audio-driven pose generation methods directly regress head pose parameter sequences from audio features~\cite{chen2020talkinghead,yi2020audio,zhang2021facial}, which does not agree with the fact that given a fixed audio sequence, different people even the same person singing the same song multiple times can produce mostly different head pose sequences as shown in Fig.~\ref{fig:head_pose_dynamics}.

We find that although the dynamics of head pose vary when the same person sings the same song multiple times, as shown in Fig.~\ref{fig:head_pose_dynamics}, the speed of head pose keeps similar in line with the rhythm of the music. Motivated by this, we propose to generate the moving speed and moving direction of the head separately, and combine them to generate head pose $p \in \mathbb{R}^6$ including Euler angles and a 3D translation vector at each time step.

\textbf{Moving Speed Generation:}
At the first stage, we use an MLP network $\varphi_{speed}$ to predict the speed of head pose parameters according to the encoded audio features at current time $t$: 

\begin{equation}
    \hat{s}_t = ABS(\varphi_{speed}(l^{pose}_t \oplus m^{pose}_t)),
\end{equation}
where $l^{pose}$ and $m^{pose}$ are the lyric and melody embedding features for pose generation, $\hat{s}_t \in \mathbb{R}^6$ is the output head speed at time step $t$. As $\hat{s}_t$ can not be negative, we apply absolute function $ABS$ to the output of $\varphi_{speed}$.

\textbf{Moving Direction Generation:} We use an LSTM network followed by a fully connected layer $\varphi_{direc}$  to generate the direction of head movements from the encoded audio features concatenated with the previous head pose and moving velocity at the last time step: 

\begin{equation}
\begin{split}
\hat{v}_{t-1} &= \hat{p}_{t-1} - \hat{p}_{t-2},\\  
o_t, c_t &= \text{LSTM}(l^{pose}_t \oplus m^{pose}_t \oplus \hat{p}_{t-1} \oplus \hat{v}_{t-1}, c_{t-1}),\\
\hat{d}_t &= tanh(\varphi_{direc}(o_t)),
\end{split}
\end{equation}
where $\hat{p}_{t-1}$, $\hat{p}_{t-2} \in \mathbb{R}^6$ indicate the generated head pose parameters represented by Euler angles and 3D translation parameters, $\hat{v}_{t-1} \in \mathbb{R}^6$ is the predicted head pose velocity, $c_{t-1}$ and $c_t$ are the cell states, $o_t$ means the output of LSTM network, $\hat{d}_t \in \mathbb{R}^6$ is the predicted moving direction, respectively at the corresponding time step.

\textbf{Head Pose Generation:} Finally, the pose $p_{t}$ at time step $t$ can be directly calculated by: $\hat{p}_t = \hat{p}_{t-1} + \hat{s}_t\times \hat{d}_t$.

\subsubsection{Eye State Generation Network}

Traditional methods usually generate only random eye blinks from audio features~\cite{zhang2021facial} or noise inputs~\cite{sinha2020identitypreserving}, ignoring some long-time eye closing phenomena in singing scenarios, \emph{e.g.}, people may close their eyes for a long time while singing the climax of the song. We decompose the generation process of eye states into random eye blinking generation and long-time eye closing state generation. Human blinks occur randomly and can be sampled from experimental predefined random distributions, but for long-time eye closing state generation, it should be learned from data.

\textbf{Random Eye Blinking Generation:} The normal blinks of human show regularity regarding the average human eye blinking rate and the average inter-blink duration~\cite{zhang2021facial}. 
Accordingly, we uniformly sample the blink interval $B_i\sim\mathcal{U}(a_i,b_i)$ and blink duration $B_d\sim\mathcal{U}(a_d, b_d)$ with the empirical  parameters $a_i=1.2s, b_i=2.0s, a_d=0.10s, b_d=0.45s$.

Then we generate the eye state of blink dynamics $\hat{e}^{blink}\in\{0,1\}$ according to $B_i$ and $B_d$.

\textbf{Long-time Eye Closing State Generation:}

We employ an MLP network $\varphi^{eye}$ to generate the eye state $\hat{e}^{long}_t$ at time step $t$:

\begin{equation}
    \hat{e}^{long}_t = \varphi^{eye}(l^{eye}_t \oplus m^{eye}_t).
\end{equation}

We combine the $\hat{e}^{blink}_t$ and $\hat{e}^{long}_t$ to get the composite dynamics of eye states $\hat{e}_t$:

\begin{equation}
    \hat{e}_{t}= \begin{cases}\hat{e}_{t}^{long}, & \text { if } \hat{e}_t^{long}>0, \\ \hat{e}_{t}^{blink}, & \text { otherwise }.\end{cases}
\end{equation}

Finally, we apply a temporal gaussian filter on $\hat{e}_t$ to get more smooth eye state dynamics. 

\subsection{Learning Objective}

We supervise our generator with the following loss functions:

\begin{equation}
L_{Reg} = L_{exp}+L_{pose}+L_{eye}+L_{att},
\end{equation}
where $L_{exp}$, $L_{pose}$ and $L_{eye}$ are the losses for facial expression, head pose, and eye states, respectively. $L_{att}$ is the loss term for pushing ATM to select useful feature channels. Each loss term is formulated as:

\begin{equation}
\begin{split}
L_{exp}&=w_1 L_{MSE}(\mathbf{f},\mathbf{\hat{f}})+w_2 L_{VEL}(\mathbf{f},\mathbf{\hat{f}}),\\
L_{pose}&=w_3 L_{MMD}(\mathbf{p},\mathbf{\hat{p}}) \\
&+ w_4 L_{L_1}(ABS(\mathbf{v}),ABS(\mathbf{\hat{v}})),\\
L_{eye}&=w_5 L_{L_1}(\mathbf{e^{long}},\mathbf{\hat{e}^{long}})\\&+ w_6 L_{MMD}(\mathbf{e^{long}},\mathbf{\hat{e}^{long}}),\\
L_{att}&= ||\mathbf{att^{exp}}||_1 + ||\mathbf{att^{pose}}||_1 + ||\mathbf{att^{eye}}||_1,\\
\end{split}
\end{equation}
where $w_1$, $w_2$, $w_3$, $w_4$, $w_5$, $w_6$ are balancing weights. $\mathbf{f}$, $\mathbf{p}$, $\mathbf{v}$, $\mathbf{e^{long}}$ are vectors containing the time serial ground truth parameters of facial expression, head pose, head moving velocity and long-time closing eye state parameters~(note that we only learn long-time closing eye dynamics from data), ranging from $t=1,2,...,T$. $\mathbf{\hat{f}}$, $\mathbf{\hat{p}}$, $\mathbf{\hat{v}}$, $\mathbf{\hat{e}^{long}}$ are the corresponding predicted vectors. $\mathbf{att^{exp}}$, $\mathbf{att^{pose}}$, $\mathbf{att^{eye}}$ are the predicted attention matrices for tasks of facial expression generation, head pose generation, and eye state generation, respectively. $ABS(\mathbf{x})$ denotes taking absolute values for each elements. We only supervise the absolute speed of generated head pose dynamics here, guiding the network to generate more rhythmical head pose dynamics aligned with music. $L_{MSE}(\mathbf{x},\mathbf{\hat{x}})=\frac{1}{T}\Vert{\mathbf{x}} - \mathbf{\hat{x}} \Vert_2^2$ is an $L^2$ norm loss term, $L_{L1}(\mathbf{x},\mathbf{\hat{x}})=\frac{1}{T}\Vert{\mathbf{x}} - \mathbf{\hat{x}} \Vert_1^1$ is an $L^1$ norm loss. $L_{VEL}(\mathbf{x}, \mathbf{\hat{x}})=\frac{1}{T-1}\sum_{t=1}^{T-1}\Vert ({x}_t-{x}_{t-1})-({\hat{x}}_t-{\hat{x}}_{t-1})\Vert_2^2$ is the velocity loss term, and $L_{MMD}$\cite{c32} is the maximum mean discrepancy loss to match all orders of statistics between the prediction and ground-truth. Here we use $\mathbf{x}$ to represent the ground-truth, while using $\mathbf{\hat{x}}$ for the predicted values. In our experiments, we empirically set $w_1=5$, $w_2=50$, $w_4=10$, $w_5=5$, and set other weights to $1.0$. 

Furthermore, in order to improve the diversity of generation results, we use an adversarial loss to fool the discriminator D, which is defined as :

\begin{equation}
\begin{split}
    L_{Adv}&=arg\mathop{\min}_{G}\mathop{\max}_{D}\mathbb{E}_{\mathbf{p,e^{long},a}}[logD(\mathbf{p,e^{long},a})]\\
    &+\mathbb{E}_{\mathbf{a},p_0}[log(1-D(G(\mathbf{a},p_0),\mathbf{a}))].
\end{split}
\end{equation}

The total loss function in training phase is:
\begin{equation}
\label{equ:final_loss}
    L=\lambda_1 L_{Reg} +\lambda_2 L_{Adv}.
\end{equation}

\section{Experiments}\label{Experiments}

\subsection{Implementation Details}

Our method is implemented with PyTorch, and all the experiments are conducted on two NVIDIA RTX 3090 GPUs. For network training, we randomly sample the frame sequence with a sliding window of 128 frames. We adopt Adam optimizer during training, with a learning rate of 0.0001 for 50 epochs. Linear learning rate decay is adopted for the last 60\% epochs. The hyperparameters in Eq.~\eqref{equ:final_loss} are $\lambda_1 = 1$ and $\lambda_2=0.1$, respectively. To get vivid and photo-realistic visualization results, we train a rendering-to-video network by following FACIAL \cite{zhang2021facial}.

\subsection{Dataset Organization}

As mentioned above, popular conventional datasets only contain talking face videos that lack expressiveness. To overcome this, we build a new dataset called SingingFace. SingingFace includes more than 600 singing videos with 6 human subjects. Our supplementary video shows the learned style of different subjects when training across all the 6 human subjects. 

\textbf{Video Collection:}
We organize our dataset by recording singing videos ourselves. Specifically, we collect the singing audio set first, then the face region of the person singing the song with music played simultaneously is recorded. Finally, we automatically align each video to the corresponding music audio using SyncNet~\cite{syncnet} to ensure audio-visual synchronization. 

\textbf{Audio Separation:}
We use a state-of-the-art audio source separation model Spleeter~\cite{spleeter2020} to extract the human voice as lyric information and the background music as melody information, respectively.

\textbf{3D Face Reconstruction:}
To automatically extract face expression parameters and head poses from a singing video, we adopt Deep3DFace~\cite{c43} to extract face parameters $[\alpha, \beta, \delta, \gamma, p]$, where $\alpha \in \mathbb{R}^{80}$, $\beta \in \mathbb{R}^{64}$, $\delta \in \mathbb{R}^{80}$ are the corresponding coefficient vectors for geometry, expression and texture. $\gamma \in \mathbb{R}^{27}$ is the spherical harmonics (SH) illumination coefficients. The 3D face pose $p=[R;t]$ is represented by rotation $R \in SO(3)$ and translation $t \in \mathbb{R}^3$. The PCA basis of geometry, texture, and expression are adopted from the Basel Face Model \cite{paysan20093d} and FaceWareHouse \cite{c3}.

\textbf{Eye State Extraction:}
We employ a state-of-the-art facial analysis system OpenFace \cite{c2} to extract action unit AU45r as the eye blink parameters. Note that we observed that the distribution of the extracted AU45r values for different people varies much, so we apply min-max normalization on AU45r for each video individually. Then we apply a time length threshold $\tau=0.5s$ to detect the short-time blinking and long-time eye closing states separately.

\textbf{Data Statistics:}
We collect over 600 Chinese and English singing videos totaling 40 hours with 30 FPS. Each video contains one person singing a whole song and the average length of videos is about 4 minutes. Each video has a stable camera location and appropriate lighting conditions. We randomly split out 90\% of the videos for training and 10\% for testing.

\subsection{Ablation Study} 

To verify the effectiveness of the key ingredients in our proposed method, \emph{i.e.}, 1) the audio separation step and two-stream audio encoder~(TSAE), 2) the attention modulator~(ATM), and 3) the head pose generation network~(PGN), we study the following variants of our method:

\begin{itemize}
    \item \textbf{Single-Stream:} no audio source separation; a single stream audio encoder is employed to encode the MFCC feature of the mixed audio wave; no ATM; and replace our PGN with an MLP network.
    
    \item \textbf{Two-Stream:} equipped with audio source separation and TSAE; no ATM; and replace our PGN with an MLP network.
    
    \item \textbf{With-ATM:} equipped with audio source separation, TSAE and ATM, and replace our PGN with an MLP network. 
    
    \item \textbf{Ours:} equipped with audio source separation, TSAE, ATM, and our proposed PGN.
    
\end{itemize}

\begin{figure}[t!]
    \centering
    \includegraphics[width=0.45\textwidth]{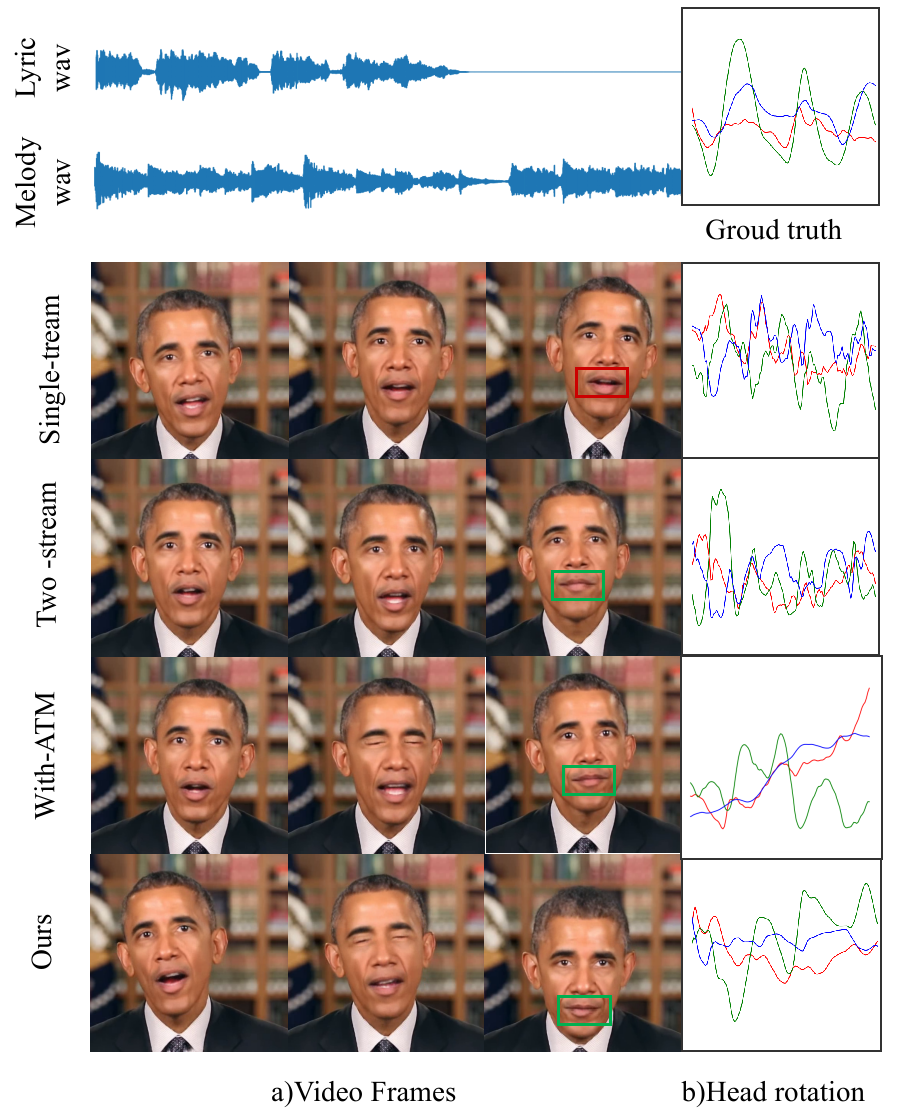}
    \caption{\textbf{Qualitative Result of Ablation Study.} It shows a) the generated frames and b) pitch~(red), yaw~(blue), and roll~(green) of head pose dynamics. In a), the mouth generated by Single-stream still keeps open during silence~(red box) while others keep closed~(green box). In b), our generated head pose dynamics are smoother than others. And the turn of dominant varying angle~(shown as green curve) occurs nearly at the same time with ground truth, meaning that our generated head dynamics have more similar rhythm to the ground truth recorded by a performer.}
    \label{fig:ablation}

\end{figure}


\begin{table}[tbp]
\centering
\caption{\textbf{Ablation Study.}}
\resizebox{\linewidth}{!}{
\begin{tabular}{lcclcclc}
\hline
\multirow{2}{*}{Method} & \multicolumn{2}{c}{Lip Sync} &  & \multicolumn{2}{c}{Pose Realism} &  & Eye Realism \\ \cline{2-3} \cline{5-6} \cline{8-8} 
                        & AVC$\uparrow$          & LMD$\downarrow$           &  & CCA$\uparrow$            & Rough$\downarrow$           &  & CCA$\uparrow$         \\ \hline
Single-Stream           & 1.17             & 1.31          &  & 0.22           & 0.24            &  & 0.06       \\
Two-Stream              & 1.73             & 1.17          &  & 0.25           & 0.18            &  & 0.07       \\
With-ATM                & 1.87             & 1.10          &  & 0.29           & 0.07            &  & 0.11        \\
Ours                    & \textbf{1.90}             & \textbf{1.08}          &  & \textbf{0.33}           & \textbf{0.06}            &  & \textbf{0.12}             \\ \hline
\end{tabular}
}
\label{tab:ablation}
\end{table}

We compare the above variants using the splitted test set of SingingFace. We evaluate the Audio-Visual Confidence~(AVC) scores proposed in \cite{syncnet}, and Landmark Distance (LMD) introduced in \cite{chen2018lip} for lip synchronization comparison. However, to the best of our knowledge, there are no clear metrics for evaluating the realism of generated head pose and eye closing dynamics for now, which is a subjective task and is an open question to the public. Following Zhang \etal \cite{zhang2021flow}, we employ Canonical Correlation Analysis~(CCA) on the generated head pose parameters sequences and eye state sequences with the ground truth and compute the Canonical Correlation as the evaluation metric for perceptual realism. Note that to emphasize the evaluation for the rhythm of the head pose dynamics that should be in line with music, we apply Canonical Correlation Analysis on the moving speed of generated head pose sequences instead of head pose parameters themselves. We also compute the second derivative based roughness~(Rough) of the generated Euler angles defined in Eq.~\eqref{equ:rough} for head motion smooth evaluation:

\begin{equation}
{\small Rough(R) = \frac{1}{T}\sum\nolimits_{t=1}^T R''(t)^2},
\label{equ:rough}
\end{equation}
where $R^{''}(t)$ denotes the second derivatives of head rotation angles at time step t. The quantitative results of ablation study are summarized in Tab.~\ref{tab:ablation}.

\textbf{Effectiveness of Two Stream Design:}~As mentioned above, lyric and melody information are entangled together in plain music waves, making it difficult to learn facial dynamics in line with the music. It's verified that, by separating human voice and background music from plain music waves and encoding the features separately, our two-stream design greatly reduces the complexity of the lip synchronization task, thus leading to a better synchronization result. As shown in Fig.~\ref{fig:ablation}, if we just learn singing facial dynamics from plain music~(Single-stream), the generated mouth movements are severely disrupted by the background music~(\emph{e.g.}, the mouth still keeps open during silence). On the contrary, the other variants that apply our two-stream design perform correctly. On the other hand, after applying source separation and our TSAE(Two-stream), all of the evaluation metrics have been improved a lot compared with Single-stream shown in Tab.\ref{tab:ablation}. This improvement can be more clearly seen in our supplementary video.

\textbf{Effectiveness of Attention-based Modulator:}~Our Attention-based Modulator automatically assigns optimal attention weights on different features at each time step for each downstream generation task. It allows our model to extract as much useful information as possible from the entangled audio features for each specific downstream task and eliminate the interference sources. This is verified from the experimental results that our ATM variant outperforms Two-stream on all of the evaluation metrics summarized in Tab.~\ref{tab:ablation}.

\textbf{Effectiveness of Pose Generation Network:}~Compared with simple MLP networks, the head pose dynamics generated by our PGN show superior perceptual results. The improvement comes from that our PGN decompose head pose sequence generation task into moving speed generation and moving direction generation. Firstly, the network is able to concentrate on the generation of moving speed which is more related to the rhythm of music, resulting in more rhythmical head pose dynamics that are in line with the music. This is verified in Tab.~\ref{tab:ablation}, that our method outperforms others a lot on the CCA metric of head pose. Then, the LSTM module for moving direction generation is able to consider not only the current audio features but also the generated head moving history, resulting in the more smooth and spontaneous head moving curves. As shown in Fig.~\ref{fig:ablation}, the visualization of pose rotation curves generated by our method~(Ours) are smooth and resemble closely the ground truth. Specifically, the turn of the dominant varying angle~(shown as green curve) of our generated head occurs nearly at the same time with ground truth. It's recommended to check our supplementary video to compare the difference more clearly.

\begin{figure}[t!]
    \centering
    \includegraphics[width=0.45\textwidth]{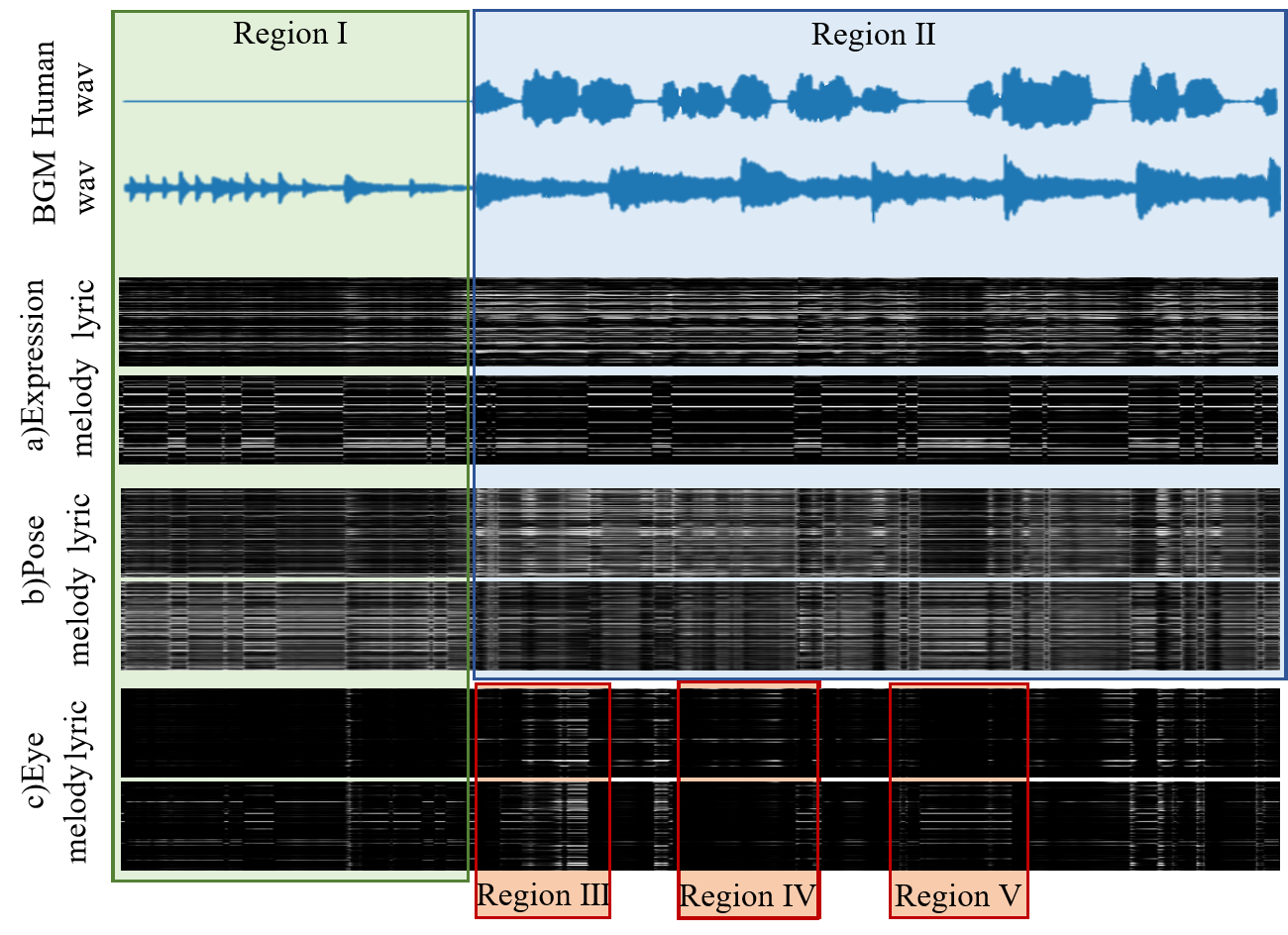}
    \caption{\textbf{Attention Weight Visualization.} The brighter white represents higher weights. The horizontal direction is along the time, and the vertical direction is along the feature dimension.}
    \label{fig:visattention}

\end{figure}

\begin{figure*}[ht!]
\centering
\resizebox{.95\textwidth}{!}{
\begin{tabular}{ccclcl}
\quad & \multirow{5}{*}{\rotatebox{90}{{\large ATVG}\cite{chen2019hierarchical}}}  & \multirow{5}{*}{\includegraphics[width=0.55\linewidth]{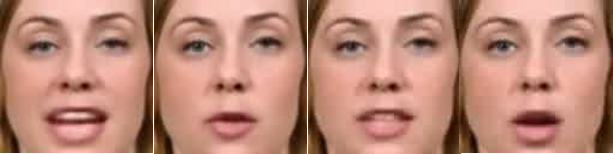}} & \multirow{5}{*}{\includegraphics[]{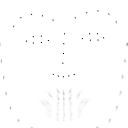}}& \multirow{5}{*}{\includegraphics[width=0.55\linewidth]{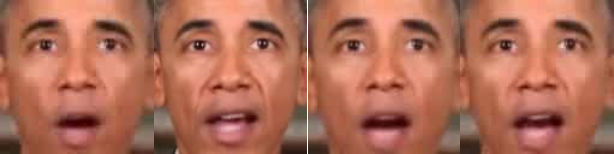}}  & \multirow{5}{*}{\includegraphics[]{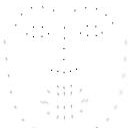}}\\
                &  &                    &          & &        \\
                &  &                  &          & &        \\
                &  &                   &          & &        \\
                &  &                   &          & &        \\
                &  &                   &          & &        \\
\quad & \multirow{5}{*}{\rotatebox{90}{{\large Yi} \textit{et al.} \cite{yi2020audio}}}  & \multirow{5}{*}{\includegraphics[width=0.55\linewidth]{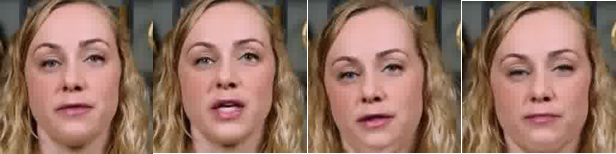}} & \multirow{5}{*}{\includegraphics[]{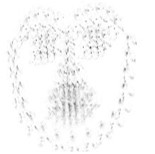}} & \multirow{5}{*}{\includegraphics[width=0.55\linewidth]{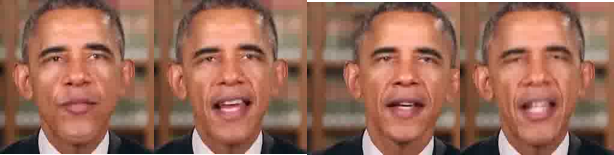}}  & \multirow{5}{*}{\includegraphics[]{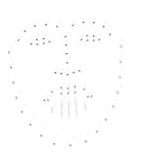}}\\
                &  &                    &          & &        \\
                &  &                  &          & &        \\
                &  &                   &          & &        \\
                &  &                   &          & &        \\
                &  &                   &          & &        \\
\quad & \multirow{5}{*}{\rotatebox{90}{{\large LSP\cite{lu2021live}}}}  & \multirow{5}{*}{\includegraphics[width=0.55\linewidth]{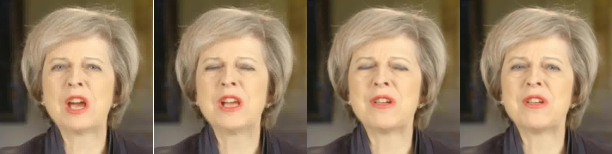}} & \multirow{5}{*}{\includegraphics[]{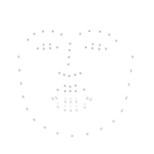}}& \multirow{5}{*}{\includegraphics[width=0.55\linewidth]{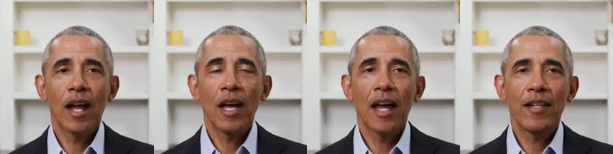}}  & \multirow{5}{*}{\includegraphics[]{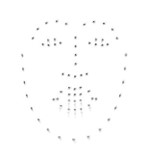}}\\ 
                &  &                    &          & &        \\
                &  &                  &          & &        \\
                &  &                   &          & &        \\
                &  &                   &          & &        \\
                &  &                   &          & &        \\

\quad & \multirow{5}{*}{\rotatebox{90}{{\large FACIAL} \cite{zhang2021facial}}} & \multirow{5}{*}{\includegraphics[width=0.55\linewidth]{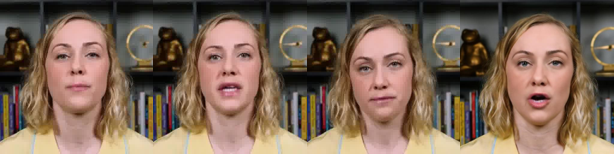}} & \multirow{5}{*}{\includegraphics[]{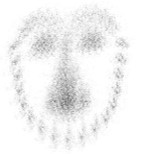}}& \multirow{5}{*}{\includegraphics[width=0.55\linewidth]{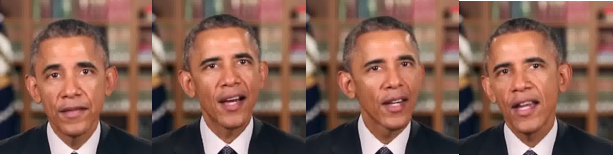}}  & \multirow{5}{*}{\includegraphics[]{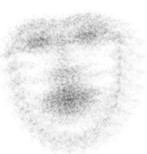}}\\ 
                &  &                    &          & &        \\
                &  &                  &          & &        \\
                &  &                   &          & &        \\
                &  &                   &          & &        \\
                &  &                   &          & &        \\
\quad & \multirow{5}{*}{\rotatebox{90}{{\large Ours}}}  & \multirow{5}{*}{\includegraphics[width=0.55\linewidth]{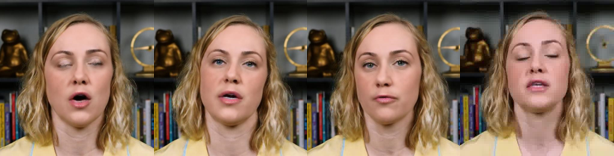}} & \multirow{5}{*}{\includegraphics[]{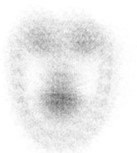}}& \multirow{5}{*}{\includegraphics[width=0.55\linewidth]{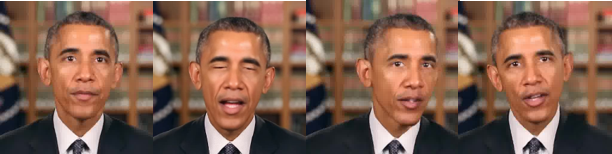}}  & \multirow{5}{*}{\includegraphics[]{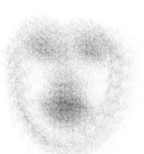}}\\ 
                &  &                    &          & &        \\
                &  &                  &          & &        \\
                &  &                   &          & &        \\
                &  &                   &          & &        \\
                &  &                   &          & &        \\
                &  &                   &          & &        \\

                &  &  a) {\Large Video frames}  & b) {\Large Tracemaps}         & c) {\Large Video frames} &  d) {\Large Tracemaps}      \\
\end{tabular}
}
\caption{\textbf{Visual Comparison with State-of-the-art Methods.} a) and c) are the generated video frames. b) and d) are the corresponding tracemaps of facial landmarks across multiple frames. From the tracemaps, we can see our method generates the most diverse head pose dynamics.}
\label{tab:total_results}
\end{figure*}

\textbf{Analysis of Attention Weight:}
To further investigate the role of the Attention Modulator~(ATM), we visualize the predicted attention weights for synthesis tasks of facial expression, head movement, and eye state. As shown in the case illustrated in Fig.~\ref{fig:visattention}, it can be observed that:
\begin{itemize}

\setlength{\itemsep}{0pt}
\setlength{\parsep}{0pt}
\setlength{\parskip}{0pt}
    \item When there is background music only and no human voice, the ATM pays more emphasis on the stream of background music (melody feature), as shown in the earlier part of the music~(See Region I). 

    \item When there is both background music and human voice, in the tasks of face expression and head pose, the ATM modulates the weights between two streams to pursue more expressive results ~(See Region II). From the numerical viewpoint, the weights of the human voice are higher than that of background music. In this case, the human voice dominates the generation of face expression and head pose. 
    \item When there is both background music and human voice, both the human voice and background music affect the long-time eye closing state, simultaneously~(See Region III) or separately~(See Regions IV and V). 
    
\end{itemize}

\subsection{Comparisons with State-of-the-art Methods}
\subsubsection{Compared State-of-the-art Methods}

Most previous state-of-the-art methods are designed for talking scenarios and trained on talking datasets such as Voxceleb2~\cite{voxceleb2} and LRS2~\cite{lrs2}. For a fair comparison, we select and retrain the methods whose training code are open to the public on our SingingFace dataset. The selected compared state-of-the-art methods are as follows:
\begin{itemize}

    \item ATVG~\cite{chen2019hierarchical} is one 2D-based cascade GAN approach to generate a talking face video that is robust to different facial characteristics, by taking an audio sequence and a target image as input.
    
    \item Yi \textit{et al.}~\cite{yi2020audio} utilize 3D face model information to synthesize photo-realistic talking face videos with personalized pose dynamics.
    
    \item LiveSpeechPortraits~(LSP)~\cite{lu2021live} presents a live system utilizing 2D landmarks to generate personalized photorealistic talking-head animation in real time.
    
    \item FACIAL~\cite{zhang2021facial} integrates implicit attribute learning to synthesize 3D face animation with realistic motions of lips, head poses, and eye blinks.

\end{itemize}

We also report the qualitative comparison results with Song2Face~\cite{iwase2020song2face}, which is the only one method designed for singing scenarios up to now to the best of our knowledge. Song2Face maps each human voice segment to facial expression and head rotation parameters, and uses an adaptive filter network to incorporate
information from neighboring frames for temporal stability. It should be note that Song2Face only models with single driving source~(plain human singing voice) as input, while ours supports multiple driving sources, \emph{e.g.} human voices and background music, and focuses on how to collaborate with different driving sources to generate more realistic head movements. In addition, eye states are dealt with as a part of facial expression for Song2Face, while ours decompose the generation of eye states as an individual generation task. Since the implementation of Song2Face is unavailable, quantitative evaluation with Song2Face is absence in this paper. It's recommended to see our supplementary video for better comparison.

\subsubsection{Qualitative Comparison}

\begin{figure}[tbp!]
\begin{tabular}{cc}
\rotatebox{90}{Song2Face~\cite{iwase2020song2face}} &\includegraphics[width=0.88\linewidth]{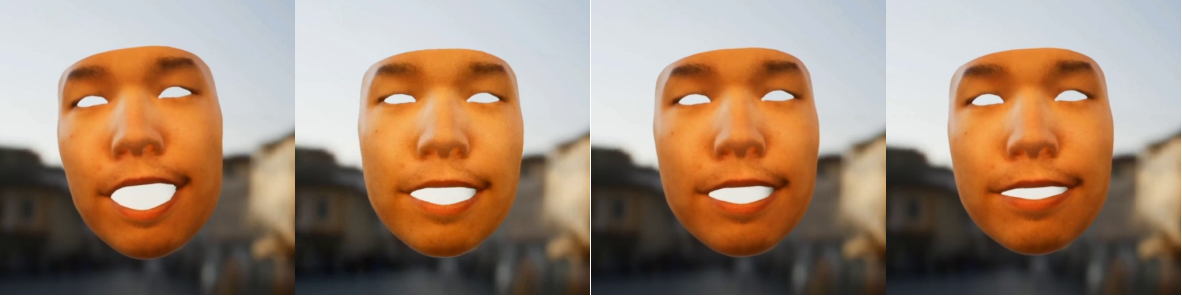}  
\\
\rotatebox{90}{\ \ \ \ \ \  Ours}      & \includegraphics[width=0.88\linewidth]{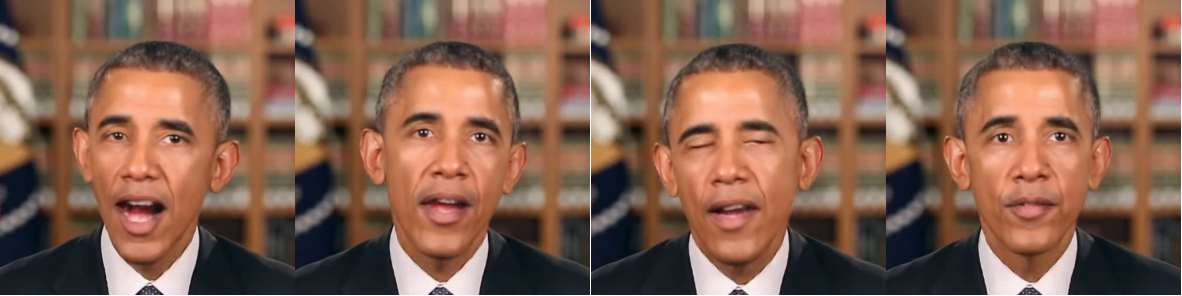} 
\end{tabular}
\caption{\textbf{Visual Comparison with Song2Face.} Our method can generate photo-realistic frames, diverse head pose and natural eye closing dynamics, which is infeasible for Song2Face~\cite{iwase2020song2face}.}
\label{tab:vssong2face}

\end{figure}

Fig.~\ref{tab:total_results} and Fig.~\ref{tab:vssong2face} shows the visual comparison with other state-of-the-art methods. We show the summary of qualitative comparison results in this section.

\textbf{Realism on Pose Dynamics:}
As shown in supplementary material, ATVG ~\cite{chen2019hierarchical} only generates talking face videos with a fully static head pose, which is against the human common sense. Yi \emph{et al.}~\cite{yi2020audio} generate photo-realistic videos but the talking faces usually show subtle movements due to the supervision pipeline. In addition, the generated head pose dynamics behave discontinuously because of the background matching trick used in \cite{yi2020audio}, which matches short-term generated head poses to one same target frame when the target frames are scarce in the target video. LiveSpeechPortraits~\cite{lu2021live} generates smooth but relatively small head movements. The generated head pose also shows a weak correlation with the rhythm of the music. FACIAL \cite{zhang2021facial} and Song2Face~\cite{iwase2020song2face} can generate more natural head pose dynamics than other compared state-of-art methods, but they still show only a few variations in head movement patterns over a long period of time. Our method can generate the most realistic head pose crediting to our pose generation method. To be specifically, for example, the head rotates quickly and dramatically during dense syllables, while slowly  during pronouncing long syllables. Readers are recommended to watch the supplementary video to see the vivid visual results more clearly.

\begin{figure}[ht!]
    \centering
    \includegraphics[width=0.4\textwidth,height=3.5cm]{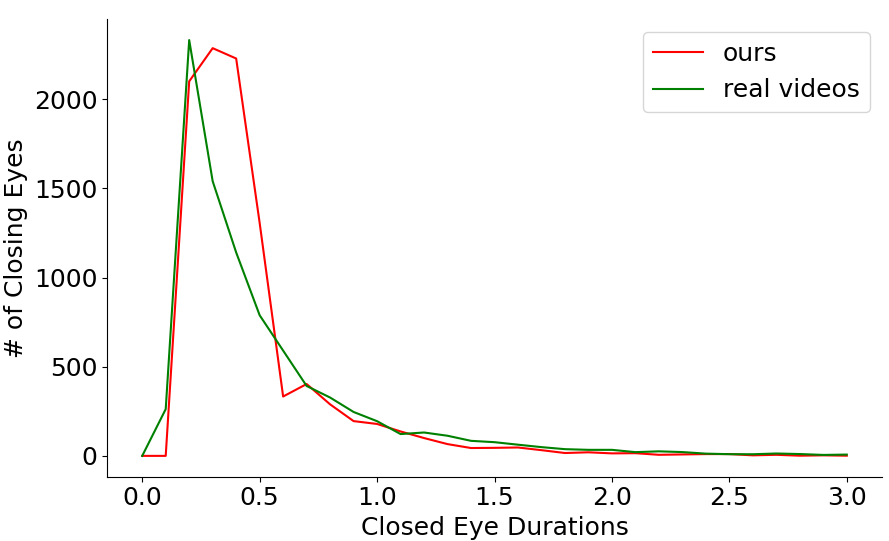}
    \caption{\textbf{Distribution of Eye Closing Duration.} Our method is able to generate realistic closed eye duration of similar distribution with the real videos. }
    \label{fig:predblinks}

\end{figure}

\textbf{Realism on Eye States:}
The generation methods for eye states between the compared methods are various. ATVG and Yi \textit{et al.} do not involve generating eye state parameters, therefore they do not produce any eye closing dynamics. For Song2Face and FACIAL, they learn random blink dynamics from data. However, Song2Face only performs well on plain human singing voice~(no background music), and FACIAL only generates open eyes during inference, failing to generate spontaneous eye closing dynamics due to the complex entanglement between short random blinks and long-time eye closing states in SingingFace dataset. LiveSpeechPortraits directly sample random blink dynamics from target video and our method synthesizes random blinks from pre-defined random distributions, both of which show realistic random blink results. Moreover, as shown in Fig.~\ref{fig:predblinks}, our method can also generate long-time eye closing dynamics (\textgreater0.5s) during voice based on the rhythm and emotion in the music, which further enhances the sense of realism.

\begin{table*}[]
\Large
\centering
\caption{\textbf{Comparisons with State-of-the-art Methods}}
\resizebox{.98\textwidth}{!}
{
\begin{tabular}{lccccccccccccc}
\hline
\multirow{2}{*}{Method} & \multicolumn{4}{c}{Mixed Wave}    &  & \multicolumn{4}{c}{Separated Wave} &  & \multirow{2}{*}{blinks/s} & \multirow{2}{*}{blink dur.(s)} & \multirow{2}{*}{CPBD$\uparrow$} \\ \cline{2-5} \cline{7-10}
                        & AVC$\uparrow$ & LMD$\downarrow$  & CCA(pose)$\uparrow$ & CCA(eye)$\uparrow$ &  & AVC$\uparrow$  & LMD$\downarrow$  & CCA(pose)$\uparrow$ & CCA(eye)$\uparrow$ &  &                           &                                &                       \\ \hline
ATVG\cite{chen2019hierarchical}                    & 0.27    & 1.46 & ---       & ---      &  & 0.34     & 1.40 & ---       & ---      &  & ---                       & ---                            & 0.11                 \\
Yi~\emph{et al.}\cite{yi2020audio}                      & 1.23    & 1.48 & 0.18      & ---      &  &    1.45  & 1.44 & 0.19      & ---      &  & ---                       & ---                            & 0.29                 \\
LiveSpeechPortraits\cite{lu2021live}     & 0.31    & 1.43 & \textbf{0.32}      & ---      &  & 0.35     & 1.42 & 0.32      & ---      &  & ---                       & ---                            & 0.20                 \\
FACIAL\cite{zhang2021facial}                  & 1.49    & 1.29 & 0.25      & 0.08     &  & 1.61     & 1.23 & 0.26      & 0.08     &  & ---                      & ---                           & 0.31                 \\
Ground Truth                      & 3.00    & ---  & ---       & ---      &  & 3.00     & ---  & ---       & ---      &  & 0.35                      & 0.23                           & 0.37                 \\
Ours                    & \textbf{1.69}    & \textbf{1.17}     & 0.26          & \textbf{0.18}         &  & \textbf{1.90}     & \textbf{1.08} & \textbf{0.33}      & \textbf{0.12}         &  & 0.38                      & 0.26                           & \textbf{0.34}                       \\ \hline
\end{tabular}
}
\label{tab:sota}
\end{table*}

\begin{table}[ht!]
\centering
\small
\caption{\textbf{Results of User Study.}}
\begin{tabular}{ccccc}
\hline
Method                & Lip  & Head & Eye  & Conformity \\ \hline
ATVG\cite{chen2019hierarchical}                  & 1.45 & 1.34 & 1.24 & 1.25       \\
Yi \etal\cite{yi2020audio}            & 2.14 & 1.70 & 1.53 & 1.78       \\
LiveSpeechPortraits\cite{lu2021live} & 1.78 & 1.80 & 2.01 & 2.27       \\
FACIAL\cite{zhang2021facial}                & 3.73 & 3.68 & 2.63 & 3.51       \\
Ours                  & \textbf{4.46} & \textbf{4.61} & \textbf{4.47} & \textbf{4.54}       \\ \hline
\end{tabular}
\label{tab:userstudy}

\end{table}

\subsubsection{Quantitative Evaluation}

We use the same test set of music with the ablation study to compare our method with state-of-the-art counterparts. To clear out the effectiveness of the audio source separation model used in our method, we report the compared results on both mixed signals and separated signals~(human voice and background music). Our method gets superior results on the most of metrics in the both cases. The results are summarized in Tab.~\ref{tab:sota}.

\textbf{Lip-sync metric:}
Similar to the ablation study, we evaluate the Landmark Distance Metric~\cite{chen2018lip} and Audio-Visual Confidence score~\cite{syncnet} to compare the lip synchronization of our method with the state-of-the-art methods. Tab.~\ref{tab:sota} shows that in both mixed and separated scenarios, our method beats all counterparts. It also shows that it is beneficial to separate the human voice from the plain music wave for mouth movement generation. Note that separated singing voice is given as input to the pre-trained lip-sync evaluation model during evaluation to ensure the pre-trained model performs correctly. 

\textbf{Pose Realism:}
In the evaluation of the realism of pose dynamics between different methods, we measure Canonical Correlation between predicted pose parameter sequences and ground-truth following with \cite{zhang2021flow}. Similarly, to emphasize the evaluation of the rhythm of the synthesized head pose sequences, we apply Canonical Correlation Analysis to the speed of the head movement. Tab.~\ref{tab:sota} shows that our method generates more realistic and rhythmic pose dynamics.

\textbf{Eye Realism:}
We measure Canonical Correlation between predicted eye state parameter sequences and ground-truth following with \cite{zhang2021flow} to evaluate the realism of long-time eye closing dynamics. For random blinking evaluation, we count the average blinking frequency (blinks/s) and intra-blink duration~(s) of generated singing face videos, and compare them with ground truth. As shown in Tab.~\ref{tab:sota}, these two statistics of our method are similar to the ground truth, falling within a reasonable range.

\textbf{Sharpness metric:}
Cumulative probability blur detection~(CPBD) is evaluated to measure the generated frame sharpness of different methods. Our implementation of renderer module generates the most sharpness facial texture according to Tab.~\ref{tab:sota}. However, as shown in our supplementary video, the generated texture of mouth region when opening mouth wide and the generated texture of eyelid when closing eyes look a little blur. The blur texture should come from the data scarcity of open mouth and closed eyes. It should be easy to improve the texture, simply by training the renderer with more data, or replacing the renderer with a few-shot face generator.

Tab.~\ref{tab:sota} shows the effectiveness of the audio source separation step for the singing face generation task, that almost all the evaluation metrics improve after decoupling human voice and background music. It also shows the superiority of our method, which generates the most realistic singing face videos and behaves better on all the evaluation metrics. 

\subsection{User Study}

We invite 15 volunteers to evaluate our method and previous works. The volunteers are a group of college students with gender balance, no previous face synthesis study experience, but are informed of the study's purpose, the standard for evaluation, and the number of compared video groups before making evaluations. The volunteers are asked to make evaluations of videos group by group. In each group, 5 synthesized videos of compared methods are shown in order. There are 5 video groups in total. During evaluating each group, the volunteers were asked to watch  all the videos of the group, then to score the videos at once based on the following criteria: 1) audio-visual synchronization, 2) natural head motion, 3) realistic eye state, 4) conformity with music. The evaluation scores include 1(very bad), 2(bad), 3(normal), 4(good), 5(very good). Finally we calculate the average scores for each method. As summarized in Tab.~\ref{tab:userstudy}, our method has the superior visual realism than previous methods.

\section{Discussion}

\textbf{Limitation and Future Work:} The proposed method achieves more expressive results against previous methods. However,  as shown in the supplementary video, under chaotic environments, our method fails like previous methods, which is because the adopted audio separator can not distinguish different human voices. On the other hand, this paper focuses on the synthesis of the head region, leaving the dynamics of the upper torso unsolved. We should note that it is more challenging to generate a realistic and expressive virtual human with dynamics of the upper torso and even the full body. This will be the direction of our future efforts. Moreover, our method just learns the implicit context from input audio, and it's indeed a interesting improvement direction to incorporate semantics from lyrics of songs.

\textbf{Ethics Statement:} 
The work itself does not uniquely raise any new ethical challenges. However, we must acknowledge that the topic of image/video synthesis has been receiving many ethical concerns. These kinds of algorithms are vulnerable to malicious use, such as potentially misused to produce misleading information or violate the portrait right. Therefore, we appeal the research community and potential users to explore the techniques responsibly. 

\appendix

\subsection*{Appendix}
Here we elaborate on more technical details of our proposed pipeline, including our Encoder, Decoder, and Discriminator. Note that we sample batches of data with frame window length of $T=128$ frames and batch size of 64 during training.

\subsection*{The Architecture of Encoder}

\textbf{The Architecture of TSAE:}
Our Two-Stream Audio Encoder (TSAE) is composed of two Single-Stream Audio Encoder (AE) with the same structure but without sharing parameters. The Single-Stream Audio Encoder is a 1D convolutional neural network with residual blocks typically used for time series classification~\cite{ismail2019deep}. The detailed architecture of our Single-Stream AE is shown in Tab.~\ref{tab:supp-AE-net}.

\begin{figure}[!tbp]
    \centering
    \includegraphics[width=0.45\textwidth]{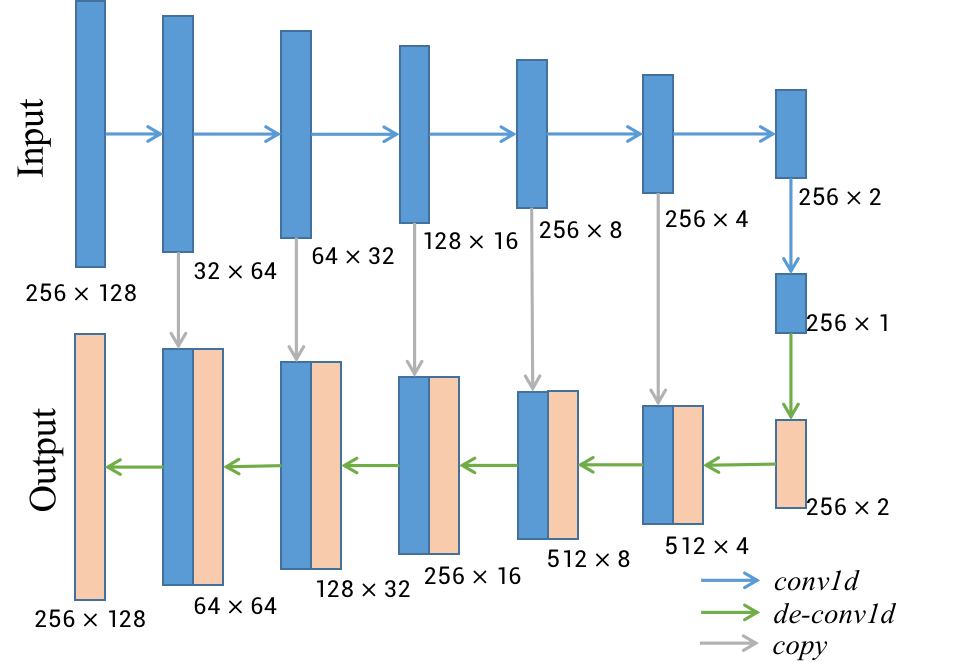}
    \caption{\textbf{The detailed U-net structure used in our Attention-based Modulator.}}
    \label{fig:unet}
\end{figure}

\textbf{The Architecture of ATM:}
Our Attention-based Modulator (ATM) is a simple unet-based 1D convolutional neural network~(CNN), taking encoded audio features $\mathbf{l} \oplus \mathbf{m} \in \mathbb{R}^{T \times 256} $ as input, computing and applying attention values on the audio features, which is similar with the channel-attention mechanism proposed in \cite{hu2018squeeze} for CNN. The U-net structure of our ATM is summarized in Fig.~\ref{fig:unet}, and the total architecture of our ATM is summarized in Tab.~\ref{tab:arch_atm}. Note that we adopt one ATM for each generation task with the same structure but without sharing parameters.

\begin{table}[!tbp]
\caption{The Architecture of Audio Encoder.}
\label{tab:supp-AE-net}
\centering
\begin{tabular}{cccc}
\hline
Type          & Downsample & Output & Activation \\ \hline
Input         & ---          & 26x39  & ---          \\ \hline
Conv1D        & False      & 32x39  & ReLU       \\
ResidualBlock & False      & 32x39  & ReLU       \\
ResidualBlock & True       & 64x20  & ReLU       \\
ResidualBlock & False      & 64x20  & ReLU       \\
ResidualBlock & True       & 128x10 & ReLU       \\
ResidualBlock & False      & 128x10 & ReLU       \\
ResidualBlock & True       & 256x5 & ReLU       \\
ResidualBlock & False      & 256x5 & ReLU       \\
ResidualBlock & True       & 512x3  & ReLU       \\
Flatten       & ---          & 1536   & ---          \\
FC            & ---          & 768    & ReLU       \\
FC            & ---          & 256    & ReLU       \\ \hline
\end{tabular}
\end{table}

\begin{table}[h!]
\caption{\textbf{Detailed Structure of ATM.} Note that we treat the last channel of input as the feature channel, so the convolution and deconvolution operations are operated over the last dimension of input, and the Multiply in the table means $\mathbf{att} \odot (\mathbf{f^v} \oplus \mathbf{f^b})$.}
\centering
\begin{tabular}{lccc}
\hline
Type     & Activation   & Output           & Output Annotation \\ \hline
Input    & ---            & $128 \times 256$ & $\mathbf{f^v} \oplus \mathbf{f^b}$  \\ \hline
Transpose     & ---            & $256 \times 128$ & ---                 \\
U-net       & ---         & $256 \times 128$ & ---                 \\
FC       & Sigmoid      & $256 \times 128$ & ---             \\
Transpose       & ---      & $128 \times 256$ & $\mathbf{att}$             \\
Multiply & ---            & $128 \times 256$ & $\mathbf{l} \oplus \mathbf{m}$      \\ \hline
\end{tabular}
\label{tab:arch_atm}
\end{table}

\subsection*{The Architecture of Decoder}

All the MLP networks in our Decoder consist of two fully connected~(FC) layers with ReLU as the activation function. The first FC layer in the MLP contains 128 nodes, while the node number of the second FC layer is determined by the task~(64 for expression generation, 6 for head pose generation, and 1 for eye state generation). The input channel of the LSTM in our Pose Generation Network (PGN) is 268~(128 for $l^{pose}_t$, 128 for $m^{pose}_t$, 6 for $p_{t-1}$ and 6 for $v_{t-1}$), and the output channel is 128.

\subsection*{The Architecture of Discriminator}
Our Discriminator is a simple CNN implemented with Conv1D, BatchNorm1D, and LeakyReLU layers. Taking concatenated audio MFCC features and generated parameters~(59 channels in total, including 26 for the human voice, 26 for background music, 6 for head pose sequence, and 1 for eye state sequence) of a time window $T=128$, the Discriminator predicts whether the input head pose and eye state sequence are real or generated. Note that we train our Discriminator using LSGAN\cite{mao2017least} for training stability. The structure of our Discriminator is summarized in Tab.~\ref{tab:discriminator}.

\begin{table}[]
\caption{The Architecture of Discriminator.}
\label{tab:discriminator}
\centering
\begin{tabular}{lcccc}
\hline
Type      & Kernel & Stride & Output          & Activation \\ \hline
Input     & ---      & ---      & $128 \times 85$ & ---          \\ \hline
Transpose & ---      & ---      & $85 \times 128$ & ---          \\
Conv1D    & 3      & 2      & $32 \times 64$  & LeakyReLU  \\
Conv1D    & 3      & 2      & $64 \times 32$  & ---          \\
BN1D      & ---      & ---      & $64 \times 32$  & LeakyReLU  \\
Conv1D    & 3      & 2      & $128 \times 16$ & ---          \\
BN1D      & ---      & ---      & $128 \times 16$ & LeakyReLU  \\
Conv1D    & 3      & 2      & $224 \times 8$  & ---          \\
BN1D      & ---      & ---      & $224 \times 8$  & LeakyReLU  \\
Conv1D    & 3      & 2      & $224 \times 8$  & LeakyReLU  \\
BN1D      & 3      & 2      & $224 \times 8$  & LeakyReLU  \\
Conv1D    & 3      & 1      & $1 \times 8$    & ---          \\ \hline
\end{tabular}
\end{table}

\subsection*{Availability of data and materials}
Song2Face has been opened to public and it's encouraged to apply for data according to the guide of official website\textsuperscript{\ref{website-appendix}}.

{

    \footnotetext[1]{https://vcg.xmu.edu.cn/datasets/singingface/index.html \label{website-appendix}
    }
}

\subsection*{Acknowledgements}

This work was supported in part by National Key R\&D Program of China (Grant No. 2021YFC3300403), National Natural Science Foundation (Grant No. 62072382), and Yango Charitable Foundation. Guo was partially supported by National Science Foundation (OAC-2007661).

\subsection*{Declaration of competing interest}

The authors have no competing interests to declare that are relevant to the
content of this article.\\

\subsection*{Electronic Supplementary Material}

A supplementary video containing the qualitative experimental results, more visual synthesized singing videos, failure cases, is available in the online version of this article.

\bibliographystyle{CVMbib}
\bibliography{refs}

\subsection*{Author biography}

\begin{biography}[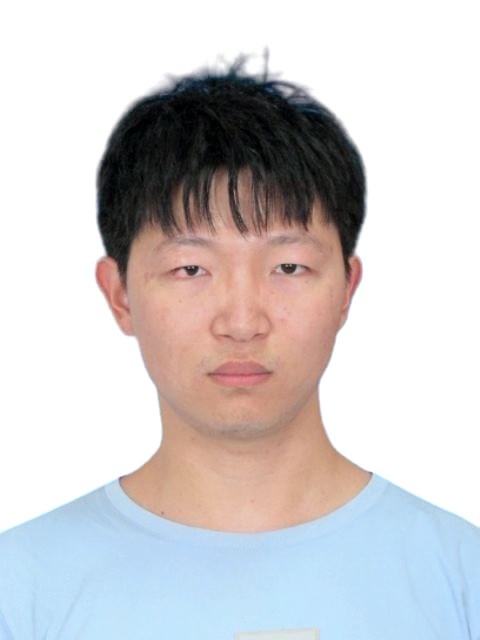]{Pengfei~Liu} is currently pursuing a master's degree at the School of Informatics, Xiamen University. He received a bachelor's degree from Xiamen University, China, in 2021. His research interests lie in the generation of digital content, especially virtual human and video conference.
\end{biography}
\vspace*{2.6em}

\begin{biography}[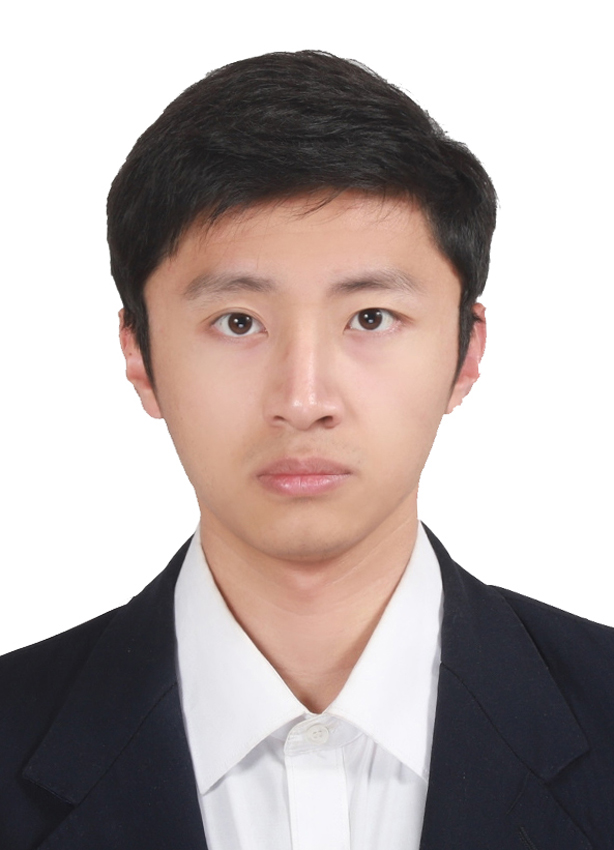]{Wenjin~Deng} is currently a postgraduate at the School of Informatics, Xiamen University. He received a bachelor's degree from Xiamen University, China, in 2020. His research interests include human pose estimation, face synthesis and avatar animation.
\end{biography}
\vspace*{2.6em}

\begin{biography}[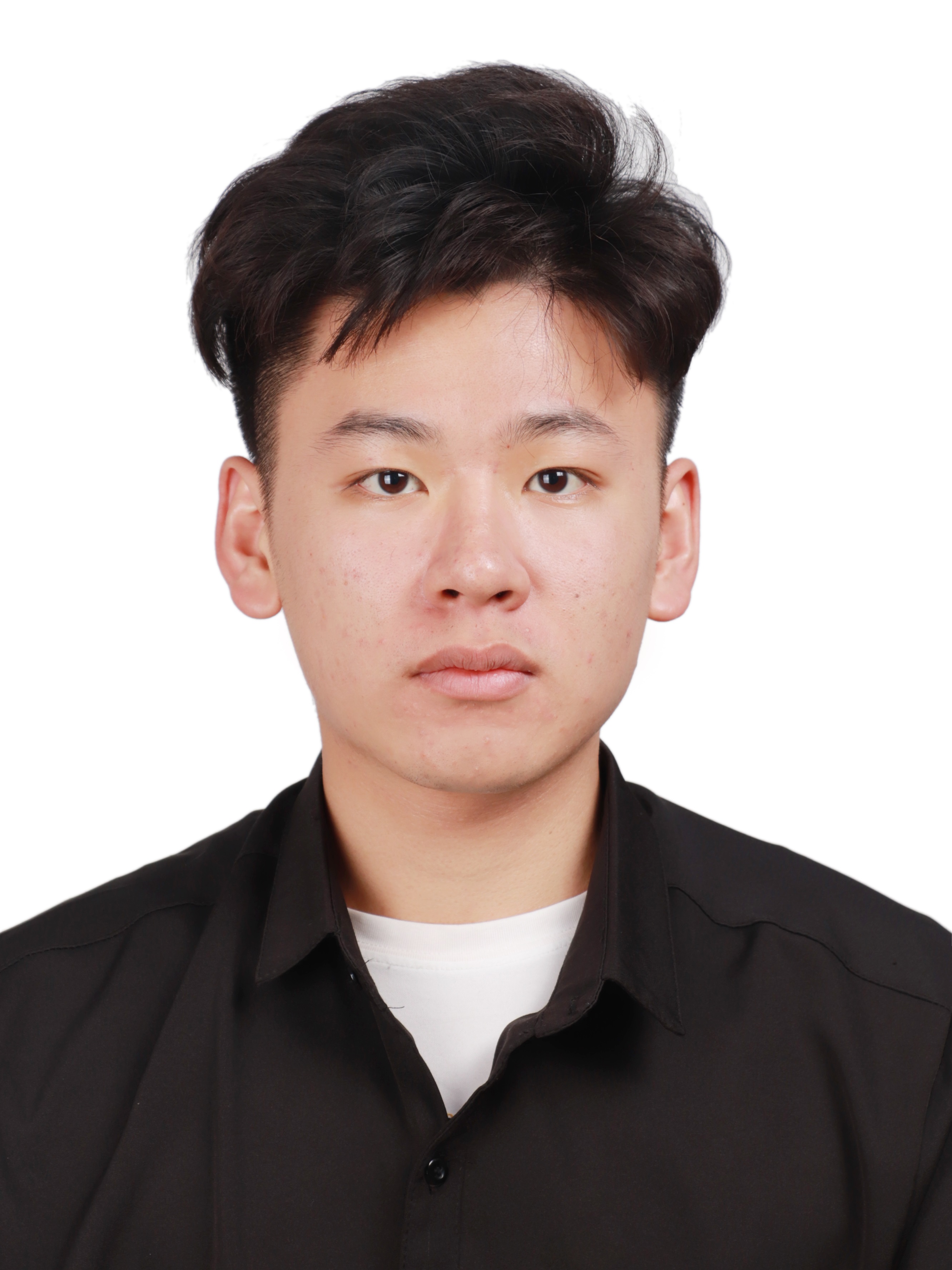]{Hengda~Li} is currently pursuing a M.S degree in School of Informatics Xiamen University, China.  He received his B.S degree from the School of Computer and Data Science Fuzhou University, China. His current research interests include face generation and face editing.
\end{biography}
\vspace*{2.6em}

\begin{biography}[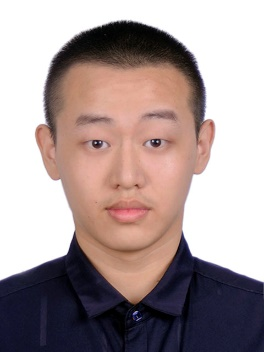]{Jintai~Wang} is pursuing a master’s degree at the Xiamen University. He received his B.Eng. degree from the School of Information, Xiamen University, in 2022. His current research interests include Neural Radiance Fields and computer vision.
\end{biography}
\vspace*{2.6em}

\begin{biography}[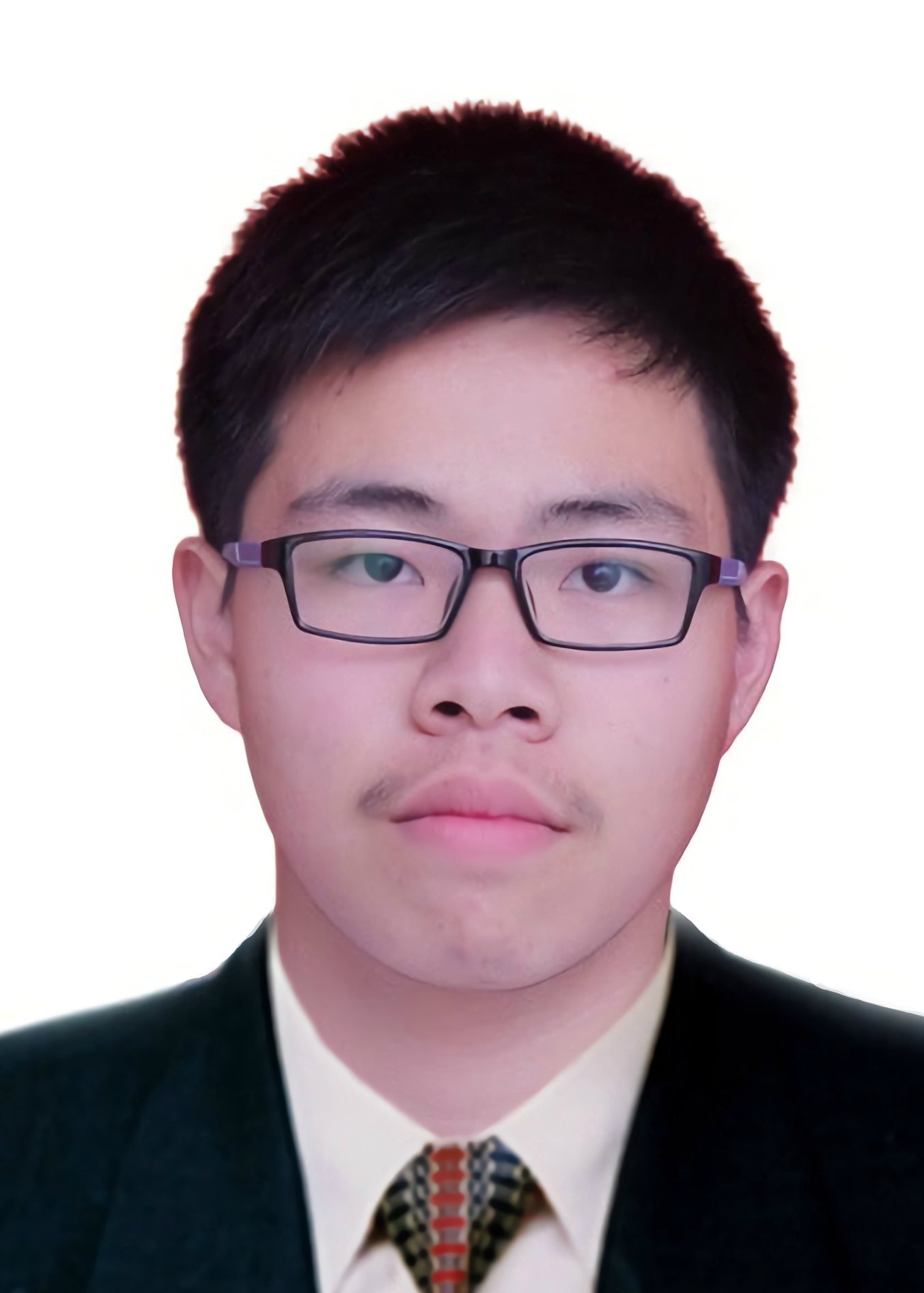]{Yinglin~Zheng} is currently pursuing a master's degree at the School of Informatics, Xiamen University. He received a bachelor's degree from Xiamen University, China, in 2020. His research interests lie in Human-related Computer Vision, especially face understanding and synthesis.
\end{biography}
\vspace*{2.6em}

\begin{biography}[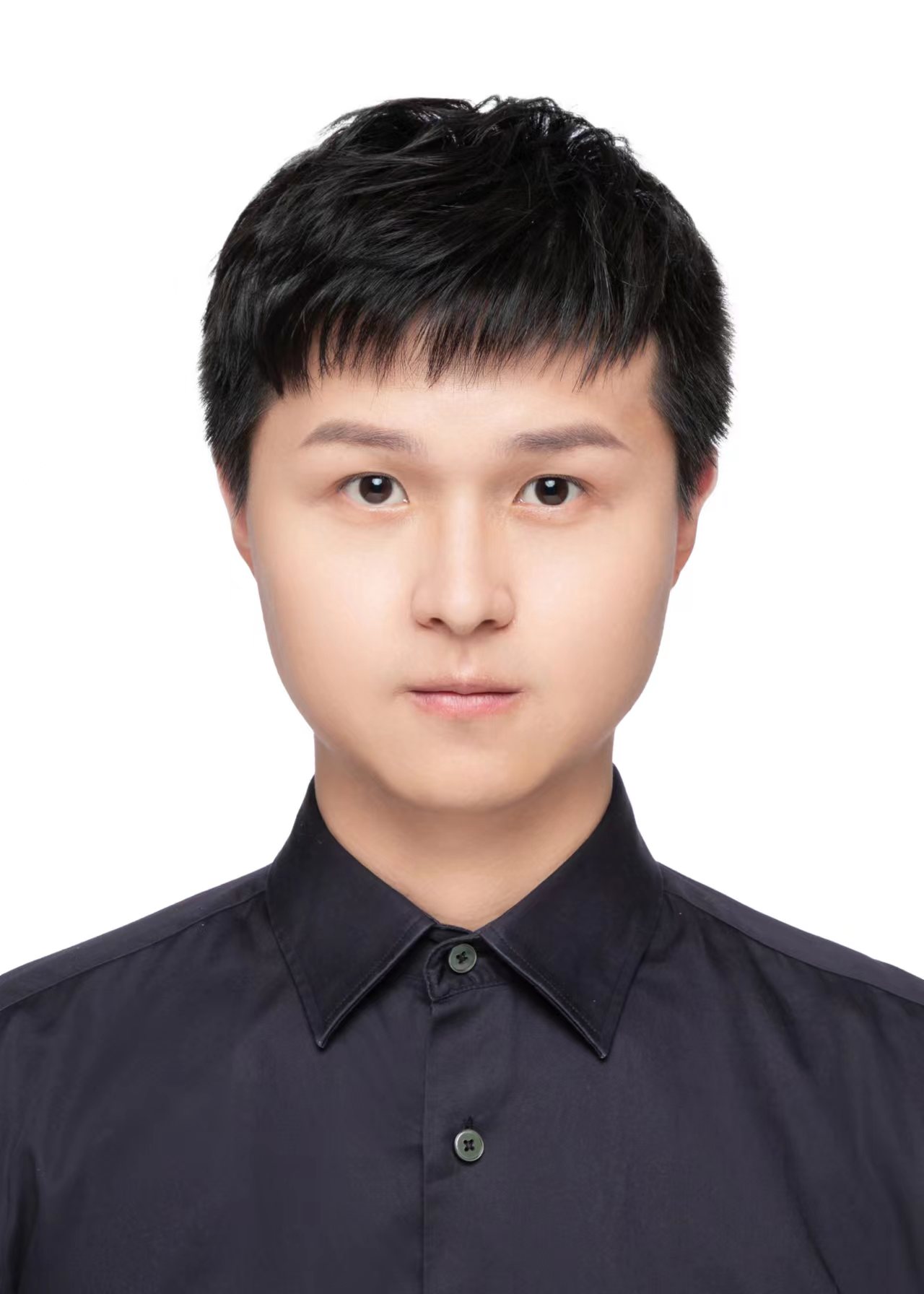]{Yiwei~Ding} is currently a postgraduate at the School of Informatics, Xiamen University. His research interests include human pose estimation, text to speech, and virtual human.
\end{biography}
\vspace*{2.6em}

\begin{biography}[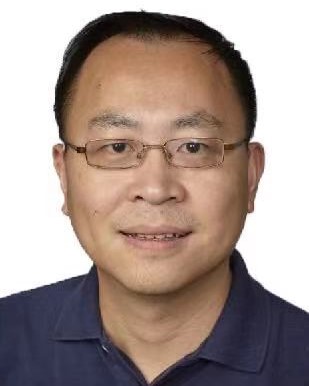]{Xiaohu~Guo} is a Full Professor of Computer Science at the University of Texas at Dallas. He received his Ph.D degree in Computer Science from Stony Brook University, and a B.S degree in Computer Science from the University of Science and Technology of China. His research interests include computer graphics, computer vision, medical imaging, and VR/AR, with an emphasis on geometric modeling and processing, as well as body and face modeling problems. He received the prestigious NSF CAREER Award in 2012. For more information, please visit \url{https://personal.utdallas.edu/~xguo/.}
\end{biography}

\vspace*{2.6em}

\begin{biography}[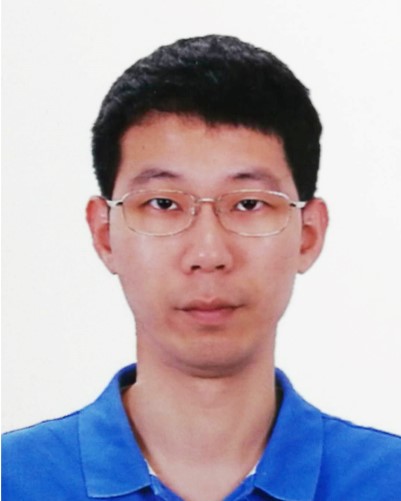]{Ming~Zeng} is currently an associate professor at the School of Informatics, Xiamen University. He was a visiting researcher at Visual Computing Group, Microsoft Research Asia (MSRA) in 2017 and 2009--2011, respectively. He received his Ph.D. degree from State Key Laboratory of CAD\&CG, Zhejiang University. His research interests include computer graphics and computer vision, especially in human-centered analysis, reconstruction, synthesis, and animation.
\end{biography}



\end{document}